\documentclass[11pt,onecolumn]{IEEEtran}
\usepackage[caption=false,font=footnotesize]{subfig}
\usepackage{graphicx}
\usepackage{booktabs}
\usepackage{amsmath}
\usepackage{amssymb}
\usepackage{midfloat}
\usepackage{bm} 
\usepackage{color, cite}
\usepackage{algorithm}
\graphicspath{{figures/}}

\newtheorem{definition}{Definition}

\newtheorem{lemma}{Lemma}

\begin{document}
\title{Improved NN-JPDAF for Joint Multiple Target Tracking and Feature Extraction}
\author{Le Zheng and Xiaodong Wang
	\thanks{Le Zheng and Xiaodong Wang are in Electrical Engineering Department, Columbia University, New York, USA, 10027, e-mail: lezheng8451@163.com, wangx@ee.columbia.edu}}
\maketitle
\begin{abstract}

Feature aided tracking can often yield improved tracking performance over the standard multiple target tracking (MTT) algorithms with only kinematic measurements. However, in many applications, the feature signal of the targets consists of sparse Fourier-domain signals. It changes quickly and nonlinearly in the time domain, and the feature measurements are corrupted by missed detections and mis-associations. These two factors make it hard to extract the feature information to be used in MTT. In this paper, we develop a feature-aided nearest neighbour joint probabilistic data association filter (NN-JPDAF) for joint MTT and feature extraction in dense target environments. To estimate the rapidly varying feature signal from incomplete and corrupted measurements, we use the atomic norm constraint to formulate the sparsity of feature signal and use the $\ell_1$-norm to formulate the sparsity of the corruption induced by mis-associations. Based on the sparse representation, the feature signal are estimated by solving a semidefinite program (SDP) which is convex. We also provide an iterative method for solving this SDP via the alternating direction method of multipliers (ADMM) where each iteration involves closed-form computation. With the estimated feature signal, re-filtering is performed to estimate the kinematic states of the targets, where the association makes use of both kinematic and feature information. Simulation results are presented to illustrate the performance of the proposed algorithm in a radar application.

\end{abstract}
\IEEEpeerreviewmaketitle

\section{Introduction}
Multiple-target tracking (MTT) is required in many applications such as surveillance, aerospace, intelligent vehicles, and monitoring of geophysical processes \cite{makris2014bayesian}. In tracking scenarios, the kinematic states of the targets are estimated from a collection of observations which may be ambiguous in origin. Thus, the difficult problem is usually the association of tracks and measurements \cite{davey2002tracking}. Algorithms such as probabilistic data association filter (PDAF) \cite{bar2009probabilistic}, joint probabilistic data association filter (JPDAF) \cite{svensson2012multitarget} and multiple hypothesis tracker (MHT) \cite{blackman2004multiple,chenouard2013multiple} have been developed to associate the targets and measurements. However, the clutters and mutiple targets can have similar locations in some challenging applications, hence algorithm based solely on kinematics is prone to mistakes \cite{ehrman2007probabilistic}.

With the advance in modern sensors, additional feature information\footnote{In this paper, we are mainly concerned with the features that are continuous. Some sensors can also provide discrete information such as number of scatterers and classification information of the targets. They are defined as attribute in \cite{drummond2003categorical}. However, the use of that kind of information is beyond the scope of this paper.} such as target sizes, shapes, and RF cross sections becomes available. On one hand, the observed feature can be used to identify the targets down to a given class \cite{smith2012robust}. On the other hand, additional information is helpful in discriminating the clutters and multiple targets, thereby improving the tracking performance. Feature aided tracking (FAT) has been gaining attention due to its significant advantage over traditional target tracking \cite{hong2005local}.

Although many algorithms have been proposed for FAT \cite{willett2001integration,song2009probabilistic,mori2014performance,ying2011feature,wang2008feature,ruan2007analytic,slocumb2005multiple}, their application is limited due to the following reason. The algorithms assume that the feature is either time-invariant or slowly-varying parameter, while the feature may change quickly and nonlinearly in practice. For example, in a radar application, the structure of specific targets has mechanical vibration or rotation, which induces a frequency modulation on the returned signal \cite{chen2006micro,luo2010micro}. As a result, the phase and amplitude of the signal change fast and periodically, which makes it hard to use the feature information in data association. Moreover, the mis-associations and missed detections lead to the corruptions in feature measurements, thereby degrading the feature extraction and target identification. These factors have made feature extractions for FAT challenging.

One characteristic that may help feature extraction is the temporal correlation of the feature signal over time, which has not been fully exploited in existing FAT approaches. However, features are extracted from a continuous string of signature signals from the same target moving over time, so how a feature changes over time is a ``feature" itself \cite{hong2005local}. In many cases, the feature signal of the target can be represented by sparse Fourier-domain signals (weighted sum of fixed frequency sinusoids) \cite{chen2006micro,stankovic2013compressive}. Another helpful fact is, for the maintained tracks, the number of mis-associations are usually small compared with the overall samples, so the sparsity also exists in the contamination induced by mis-association.

With the development of sparse signal representation and later the compressed sensing (CS) theory \cite{donoho2006compressed,eldar2012compressed}, which studies the recovery of a sparse signal from a number of linear measurements much less than its ambient dimension, sparse methods have been developed for frequency recovery \cite{stankovic2013compressive} and denoising \cite{jokanovic2015reduced, studer2012recovery}. However, CS algorithms assume the frequencies of interest lie on a fixed grid of the frequency domain because the CS focuses on signals that can be sparsely represented under a finite discrete dictionary \cite{yang2014exact}. As most feature signals are usually specified by parameters in a continuous domain, the discretization usually results in model mismatch and degradation in recovery \cite{chi2011sensitivity}.

To overcome grid mismatch of traditional sparsity-based methods, we apply the recently developed mathematical theory of continuous sparse recovery for feature extraction \cite{candes2013super,candes2014towards,tang2013compressed}. In \cite{candes2013super,candes2014towards}, the authors treat the complete data case and show that the frequencies can be exactly recovered via convex optimization once the separations between the frequencies are larger than certain threshold. In \cite{tang2013compressed,bhaskar2013atomic}, the result is extended to the problem of continuous frequency recovery from incomplete data based on the atomic norm minimization. Super resolution based on atomic norm has many applications including direction of arrival estimation \cite{tan2014direction}, channel estimation \cite{chi2015compressive} and line spectral estimation \cite{bhaskar2013atomic,tang2014robust}.

In this paper, we propose an algorithm for joint multiple target tracking and feature extraction. We consider the case when the feature signal of the target is a sparse Fourier-domain signal. In each batch of the tracker, the nearest neighbor JPDAF (NN-JPDAF) is firstly used for a rough estimate of kinematic states and extraction of feature measurements. To estimate the feature signal from incomplete and corrupted measurements, we use the atomic norm constraint to formulate the sparsity of feature and use the $\ell_1$-norm to formulate the sparsity of the corruption induced by mis-associations. The feature signal can be estimated by solving a semidefinite program (SDP) which is convex. To improve the efficiency of the algorithm, we also provide an iterative algorithm based on the alternating direction method of multipliers (ADMM) \cite{boyd2011distributed} where each iteration involves closed-form computation. Given the estimation of feature signal, a re-filtering is performed to obtain a more precise estimate of the kinematic states.

The remainder of the paper is organized as follows. Section II introduces the signal model, formulates the problem and reviews the original NN-JPDAF algorithm. In Section III, we develop the feature-aided tracking algorithm and present the sparse formulation for feature estimation. In Section IV, we present the way to solve the sparsity-based optimization for feature extraction and a summary of the proposed algorithm. Simulation results are presented in Section V, and the factors that influence the performance of the proposed algorithm are analyzed. Section VI concludes the paper.

\section{Background}
\subsection{Kinematic Model}
Consider the problem of tracking $M$ moving targets, where the $m$-th target is modeled by the following state-space equations
\begin{eqnarray}
\label{eq:x}
\bm x_m^{\mathbb K} (t + 1) &=& {\bm F_m}(t)\bm x_m^{\mathbb K} (t) + {\bm \nu_m^{\mathbb K}}(t), \\
\label{eq:y}
\bm y_m^{\mathbb K} (t) &=& \bm H_m(t)\bm x_m^{\mathbb K} (t) +\bm w_m^{\mathbb K}(t),
\end{eqnarray}
for $t=0,1,\cdots,N-1$ and $m = 1,2,...,M$. $\bm x_m^{\mathbb K}(t)$ and $\bm y_m^{\mathbb K}(t)$ are the state vector and measurement vector of the $m$-th target at time $t$ respectively, with superscript $\mathbb K$ emphasizing the kinematic characteristics. $\bm F_m(t)$ and $\bm H_m(t)$ are known model matrices. $\bm \nu_m^{\mathbb K}(t)$ and $\bm w_m^{\mathbb K}(t)$ are noise vectors, which are assumed to be zero-mean independent and identically distributed Gaussian processes with known covariances:
\begin{eqnarray}
\label{eq:noise}
\mathbb{E}\left[\bm \nu_m^{\mathbb K}(t) \bm \nu_m^{\mathbb K}(t)^T \right] &=& \bm Q_m(t), \\
\mathbb{E}\left[\bm w_m^{\mathbb K}(t) \bm w_m^{\mathbb K}(t)^T\right] &=& \bm R_m(t).
\end{eqnarray}

Suppose that $n_t$ measurements are obtained at time $t$. In a cluttered environment, $n_t$ is not necessarily equal to $M$ and it may be difficult to distinguish whether a measurement originated from a target or from clutter. Denote $\bm z_r^{\mathbb K}(t)$ as the $r$-th kinematic measurement at time $t$, then
\begin{equation}
\label{eq:pz}
\bm z_r^{\mathbb K}(t) = \left\{ \begin{gathered}
\bm y_m^{\mathbb K}(t),{\text{ if }} \bm z_r^{\mathbb K}(t) \text{ is from target } m,\hfill \\
\bm z_r^{{\mathbb K},{\cal C}} (t),{\text{ if }} \bm z_r^{\mathbb K}(t) {\text{ is from clutter}}, \hfill \\
\end{gathered}  \right.
\end{equation}
for $r=1,2,...,n_t$, where the measurement $\bm z_r^{{\mathbb K},{\cal C}} (t)$ is assumed to be uniformly distributed in the surveillance region. The number of clutters follows the Poisson distribution. We denote the kinematic measurements at time $t$ and over time steps $[0,1,...,d]$ as $\bm Z^{\mathbb K}(t)=\{\bm z_r^{{\mathbb K}}(t) : 1 \leq r \leq n_t\}$ and $\bm Z^{\mathbb K,d}= \{\bm Z^{\mathbb K}(d): 0 \leq t \leq d \}$, respectively.

\subsection{Feature Model}

We consider the feature as complex signals whose spectra consists of discrete spikes with unknown locations in the normalized interval $[0,1]$. Let $x_m^{\mathbb F}(t)$ be the feature signal of the $m$-th target at time $t$ with
\begin{eqnarray}
x_m^{\mathbb F}(t) = \sum\limits_{k = 1}^{K_m} {{c_m(k)}{e^{i2\pi f_m(k) t + i\phi_m(k)}}} ,t = 0,1,...,N - 1,
\end{eqnarray}
where $K_m$ is the number of frequency components; $c_m(k)>0$, $f_m(k)$ and $\phi_m(k)$ are the magnitude, frequency and phase of the $k$-th spike in spectrum, respectively. We denote the feature signal of the $m$-th target over $N$ time steps as $\bm x_m^{\mathbb F} = \left[ x_m^{\mathbb F}(0), x_m^{\mathbb F}(1), ..., x_m^{\mathbb F}(N-1) \right]^T$. Denote $z_r^{\mathbb F}(t)$ as the $r$-th feature measurement at time $t$, then we have
\begin{eqnarray}
\label{eq:zf}
z_r^{\mathbb F}(t) = \left\{ \begin{array}{l}
x_m^{\mathbb F}(t)+w_m^{\mathbb F}(t),{\text{ if } z_r^{\mathbb F}(t) \text{ is from target }m,}\\
z_r^{\mathbb F,{\cal C}}(t),{\text{ if } z_r^{\mathbb F}(t) \text{ is from clutter},}
\end{array} \right.
\end{eqnarray}
for $r=1,2,...,n_t$, where $w_m^{\mathbb F}(t)$ is the feature measurement noise, $z_r^{\mathbb F,{\cal C}}(t)$ denotes the feature measurement generated by clutter which is usually different from that generated by a target. We assume the variance of the feature measurement noise is $\mathbb{E}\left[\|w_m^{\mathbb F}(t)\|_2^2\right] = \sigma_m^2$.  We denote the feature measurements at time $t$ and over time steps $[0,1,...,d]$ as $\bm Z^{\mathbb F}(t)=\{z_r^{{\mathbb F}}(t) : 1 \leq r \leq n_t\}$ and $\bm Z^{\mathbb F,d}= \{\bm Z^{\mathbb F}(d): 0 \leq t \leq d \}$, respectively.

\subsection{A Motivating Example}

To motivate our work by an application, we introduce an example in radar application. Suppose we are tracking viberating targets that are modeled by vibrating scatterers. The reflected signal varies periodically. Let $\bm x_m^{\mathbb K}(t)=[x_m^{\mathbb K}(t,1),x_m^{\mathbb K}(t,2)]^T$ be the kinematic state of the $m$-th target, where $x_m^{\mathbb K}(t,1)$ is the range from the radar to the $m$-th target and $x_m^{\mathbb K}(t,2)$ is the radial velocity, which are given respectively by \cite{chen2006micro}
\begin{eqnarray}
\label{eq:range}
x_m^{\mathbb K}(t,1) &=& x_m^{\mathbb K}(0,1) + (t-1)\Delta t x_m^{\mathbb K}(0,2) + \varrho_m {\rm sin}(2 \pi f_m (t-1)\Delta t),\\
\label{eq:velocity}
x_m^{\mathbb K}(t,2) &=& x_m^{\mathbb K}(0,2) + \varrho_m 2 \pi f_m {\rm cos}(2 \pi f_m (t-1) \Delta t),
\end{eqnarray}
where $\varrho_m$ is the vibration magnitude depending on the azimuth and elevation angle of the vibration direction; $f_m$ is the frequency of the vibration.

The recieved signal from the target is regarded as the feature. Specifically, for the $m$-th target, we have
\begin{eqnarray}
	\label{eq:signal}
	x_m^{\mathbb F}(t) &=& b_m \exp \left(\frac{i4 \pi}{\xi} x_m^{\mathbb K}(t,1) + i\phi_m + i\phi_0 \right) \nonumber \\
	&=& b_m \exp \left(\frac{i4\pi}{\xi} \left(x_m^{\mathbb K}(0,1) + (t-1)\Delta t x_m^{\mathbb K}(0,2) + \varrho_m {\rm sin}(2 \pi f_m (t-1)\Delta t) \right) + i \phi_m + i \phi_0\right),
\end{eqnarray}
where $\xi$ is the wave length of the radar waveform; $\phi_0$ is the initial phase of the waveform; $b_m$ is the strength of the $m$-th target and $\phi_m$ is the phase shift of the $m$-th target.

\begin{figure}[htbp]
	\centering
	\subfloat[][]{\includegraphics[width=3.2in]{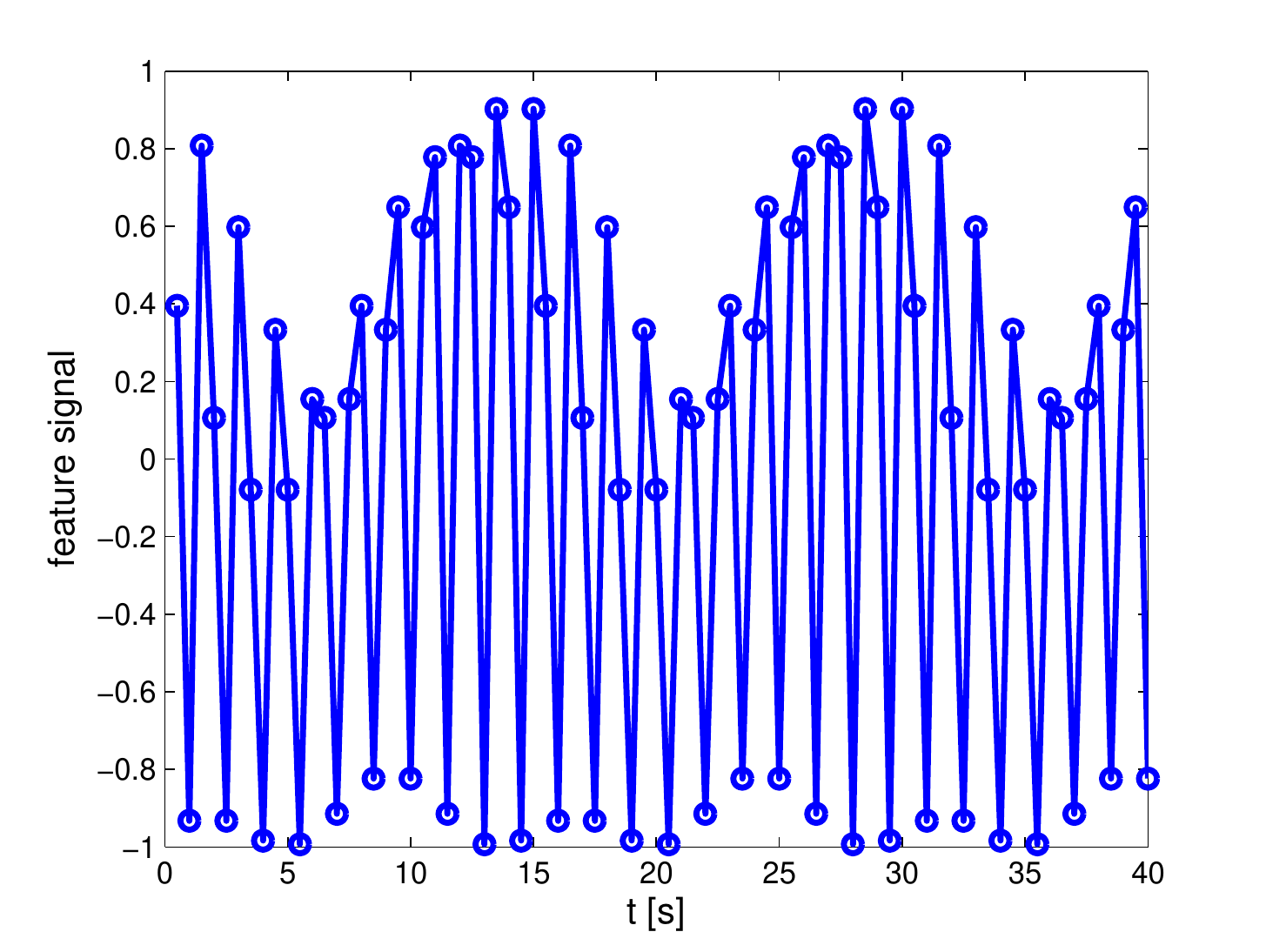}}
	\subfloat[][]{\includegraphics[width=3.2in]{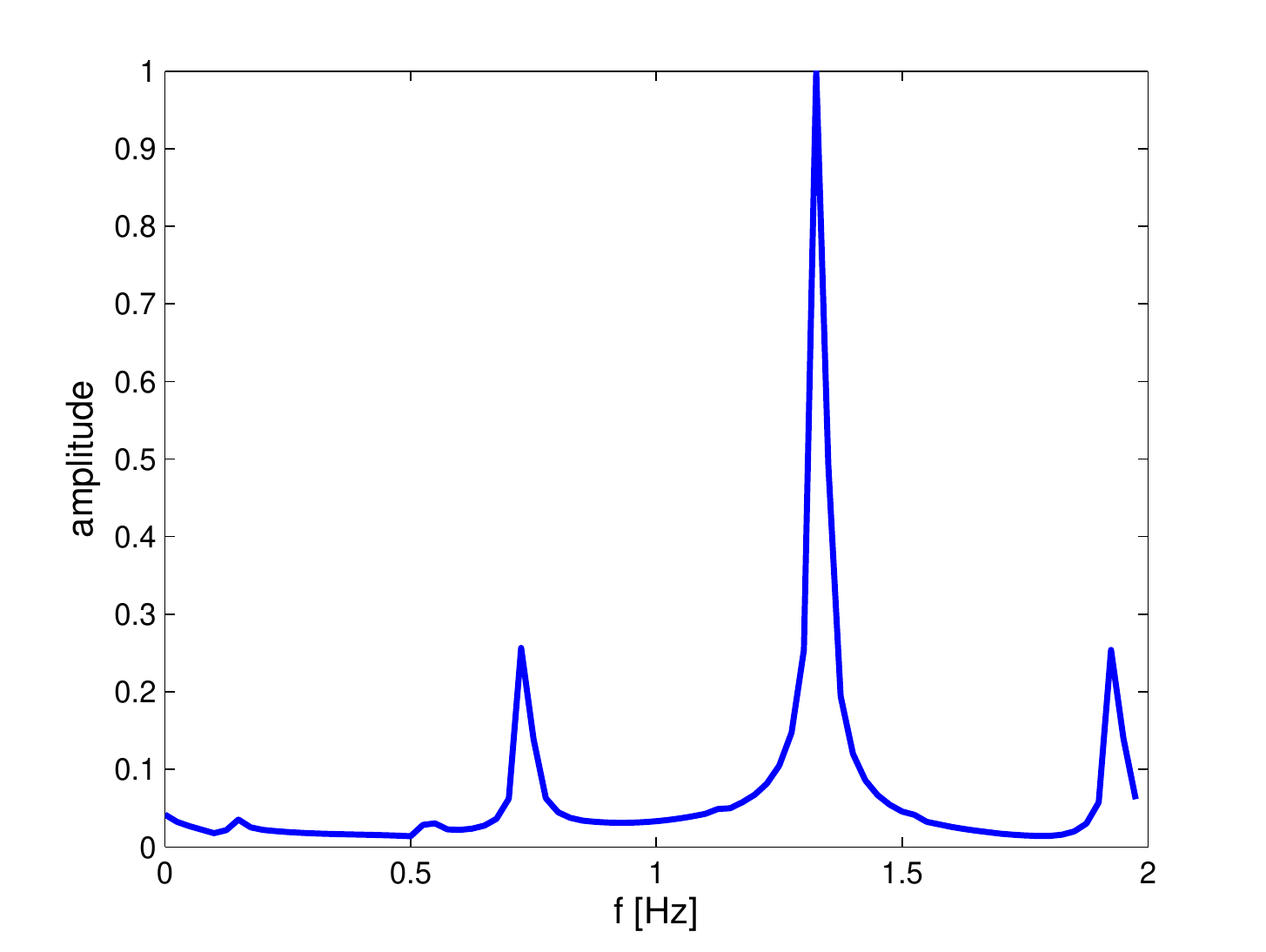}}
	\caption{(a) Plots of real parts of the signal $x_m^{\mathbb F}(t)$. (b) Frequency representation of signal reflected from the vibrating scatterer.}
	\label{fig:signal}
\end{figure}

Fig. \ref{fig:signal} shows the signal $x_m^{\mathbb F}(t)$ and its frequency spectrum with $f_m=0.6$Hz. As can be seen from the figure, the signal is sparse in the frequency domain and the spectrum consists of pairs of spectral lines around the center frequency. The vibration frequency of the scatterer can be observed by deviations of the frequency spectrum from the center frequency, and the amplitude of each frequency component can be further expressed by the Bessel function \cite{chen2006micro}. Hence, the vibration frequency can be identified by the distance between the strongest frequency component and the second strongest frequency component. In practice, the vibration frequency of different types of targets are usually different, so it can be used as an important feature for target identification.

To extract the viberation frequency of the $m$-th target, we need to have the feature measurement generated by the $m$-th target at each time, i.e., $x_m^\mathbb{F}(t)+w_m(t)$ for $t=0,1,...,N-1$. However, due to the interference of false alarms and multiple targets, there are multiple measurements at each time and the assignment between the targets and the measurements is unknown. Hence, it is not possible to directly estimate the frequencies from the feature measurements. Without the feature signal, it is also not possible to use the feature information to assist tracking.

\subsection{Outline of NN-JPDAF}
\label{sec:NN}
One popular single-scan maximum-likelihood (ML) data association technique is the nearest-neighbor joint probabilistic data association filter (NN-JPDAF), where ``nearest-neighbor" refers to how measurements are assigned to established tracks \cite{crouse2013advances}. Due to its high efficiency and good association performance, the NN-JPDAF has been widely used in radar systems for multiple target tracking \cite{bar2009probabilistic}. We define the kinematic state estimation and state estimate covariance at time $t$ as $\bm{\hat x}_m^{\mathbb K}(t|t)$ and $\bm P_m(t|t)$, respectively. Starting with the filtering result at time $t-1$, the algorithm makes the following Kalman filter prediction:
\begin{eqnarray}
	\label{eq:xpredic}
	\bm{\hat x}_m^{\mathbb K}(t|t-1) &=& \bm F_m(t-1) \bm{\hat x}_m^{\mathbb K}(t-1|t-1), \\
	\label{eq:zpredic}
	\bm{\hat z}_m^{\mathbb K}(t|t-1) &=& \bm H_m(t) \bm{\hat x}_m^{\mathbb K}(t|t-1), \\
	\label{eq:Ppredic}
	\bm P_m(t|t-1) &=& \bm F_m(t-1) \bm P_m(t-1|t-1) \bm F_m(t-1)^T+ \bm Q_m(t-1),\\
	\label{eq:Spredic}
	\bm S_m(t) &=& \bm H_m(t) \bm P_m(t|t-1) \bm H_m(t)^T + \bm R_m(t),
\end{eqnarray}
where $\bm{\hat x}_m^{\mathbb K}(t|t-1)$ is the predicted kinematic state; $\bm P_m(t|t-1)$ is the prediction covariance; $\bm S_m(t)$ is the innovation covariance.

To build the relation between tracks and measurements, a posteriori probability $\beta_{m,r}^{\mathbb K}(t)$ is computed, which is the probability that measurement $r$ originates from target $m$. Let $\chi_{m}(t)$ denote the index of measurement that is associated to track $m$, then the likelihood of $\bm z_r^{\mathbb K}(t)$ conditioned on event $\chi_{m}(t)=r$ is
\begin{eqnarray}
	\label{eq:likelihoodK}
	C_{m,r}^{\mathbb K}(t) &=& p\left(\bm z_r^{\mathbb K}(t)| \chi_{m}(t)=r, \bm Z^{\mathbb K,t-1} \right) \nonumber \\
	&=& \frac{1}{\sqrt{|2 \pi \bm S_m(t)|}} \exp\left(-\frac{1}{2}\bm v_{m,r}^{\mathbb K}(t)^T \bm S_m(t)^{-1} \bm v_{m,r}^{\mathbb K}(t) \right),
\end{eqnarray}
where $\bm v_{m,r}^{\mathbb K}(t)=\bm z_r^{\mathbb K}(t) - \bm{\hat z}_m^{\mathbb K}(t|t-1)$ is the innovation generated from $\bm z_r^{\mathbb K}(t)$ for target $m$. According to \cite{blair1999nnjpda}, the probability of the event $\chi_{m}(t)=r$ conditioned on the past measurements can be approximated by
\begin{eqnarray}
	\label{eq:beta}
	\beta_{m,r}^{\mathbb K}(t) &=&  P(\chi_{m}(t) = r | \bm Z^{\mathbb K,t}) \nonumber \\
	&\approx& \frac{C_{m,r}^{\mathbb K}(t)}{D_m^{\mathbb K}(t)+E_r^{\mathbb K}(t)-C_{m,r}^{\mathbb K}(t)+B},
\end{eqnarray}
where 
\begin{eqnarray}
	\label{eq:D}
	D_m^{\mathbb K}(t)=\sum_{r=1}^{n_t}{C_{m,r}^{\mathbb K}(t)},\\
	\label{eq:E}
	E_r^{\mathbb K}(t) = \sum_{m=1}^{M}{C_{m,r}^{\mathbb K}(t)},
\end{eqnarray}
and the constant $B$ is included to account for the nonunity probability of detection and the presence of clutters and false alarms \cite{muvsicki2007efficient}. If there is no measurement generated by track $m$, then $\chi_{m}(t)=0$.

In the NN-JPDAF, at most one measurement can be assigned to one track. The measurement-to-track assignment that gives the largest $\beta_{m,r}^{\mathbb K}(t)$ is made, i.e., $\max_{m,r} \beta_{m,r}^{\mathbb K}(t)$. Following this, all candidate associations that include either measurement $r$ or track $m$ are removed from consideration. Using the remaining $M-1$ tracks and $n_t-1$ measurements, the probabilities of the association events are computed and the measurement-to-track association that gives the new largest $\beta_{m,r}^{\mathbb K}(t)$ is made. This process continues until all measurements have been assigned to tracks or all tracks have been assigned to measurements or $\beta_{m,r}^{\mathbb K}(t)<\eta$ for all the remaining track and measurement pairs where $\eta$ is a threshold.

Let $\hat \chi_{m}(t)$ be the estimated measurement index for the $m$-th track at time $t$. For $\hat \chi_{m}(t)>0$, the NN-JPDAF uses $\bm{z}_{\hat \chi_{m}(t)}^{\mathbb K}(t)$ as if it were from the target of interest and performs the Kalman filter update:
\begin{eqnarray}
	\label{eq:xupda}
	\bm{\hat x}_m^{\mathbb K}(t|t) &=& \bm{\hat x}_m^{\mathbb K}(t|t-1) + \bm W_m(t) \bm{\tilde v}_m^{\mathbb K}(t),\\
	\label{eq:Pupda}
	\bm P_m(t|t) &=& \bm P_m(t|t-1) - \bm W_m(t) \bm S_m(t) \bm W_m(t)^T,
\end{eqnarray}
where
\begin{eqnarray}
	\label{eq:innovation}
	\bm{\tilde v}_m^{\mathbb K}(t) &=& \bm{z}_{\hat \chi_{m}(t)}^{\mathbb K}(t) - \bm{\hat z}_m^{\mathbb K}(t|t-1),\\
	\label{eq:gain}
	\bm W_m(t) &=& \bm P_m(t|t-1) \bm H_m(t)^T \bm S_m(t)^{-1}.
\end{eqnarray}
If $\hat \chi_{m}(t)=0$, the track is updated as
\begin{eqnarray}
	\label{eq:xupda2}
	\bm{\hat x}_m^{\mathbb K}(t|t) &=& \bm{\hat x}_m^{\mathbb K}(t|t-1),\\
	\label{eq:Pupda2}
	\bm P_m(t|t) &=& \bm P_m(t|t-1).
\end{eqnarray}

The original NN-JPDAF only uses the kinematic measurements for data association. If in addition there are feature measurements and the feature can be modeled by a linear dynamic process (or approximated as linear dynamic process without a major loss of information), we can directly incorporate it as an extra dimension of the state vector, so the NN-JPDAF can track both kinematic and feature states. However, the feature signal may vary rapidly and nonlinearly in practice. As a result, it is hard to estimate the feature via filtering technique. In the next section, we will introduce a sparsity-based algorithm to estimate the feature signal which is then combined with NN-JPDAF to enhance the tracking performance.

\section{Feature-Aided Tracking}

In this section, we propose a sparsity-based method to estimate the feature signal of the targets. The extracted feature can be fed back to the tracker to improve the association performance, thereby improving the tracking accuracy as well. According to the model of the previous section, the kinematic states $\bm x_m^\mathbb{K}(t)$ and feature signal $x_m^\mathbb{F}(t)$ are connected with the assignment variable $\chi_m(t)$. Specifically, if $\chi_m(t) \ne 0$, then
\begin{eqnarray}
\bm z_{\chi_m(t)}^\mathbb{K}(t) &=& \bm H_m(t) \bm x_m^\mathbb{K}(t) + \bm w_m^\mathbb{K}(t), \\
z_{\chi_m(t)}^\mathbb{F}(t) &=& x_m^\mathbb{F}(t) + w_m^\mathbb{F}(t). 
\end{eqnarray}
Obviously, the kinematic measurements and feature measurements are connected by the assignment variables. If the feature signals at each time step are known, we can modify the association probability with the feature information and obtain better association performance. Specifically, the likelihood of the feature measurement at time $t$ is
\begin{eqnarray}
\label{eq:likelihoodF}
C_{m,r}^{\mathbb F}(t) = p\left( z_r^{\mathbb F}(t)|{x}_m^{\mathbb F}(t), \chi_{m}(t) = r \right) = \frac{1}{\sqrt{2 \pi \sigma_m^2}} \exp\left(-\frac{|v_{m,r}^{\mathbb F}(t)|^2}{2 \sigma_m^2} \right),
\end{eqnarray}
where $v_{m,r}^{\mathbb F}(t)=z_r^{\mathbb F}(t)- x_m^{\mathbb F}(t)$. The joint likelihood is then given by
\begin{eqnarray}
\label{eq:likelihoodJ}
C_{m,r}^{\mathbb K,\mathbb F}(t)&=&p\left( \bm z_r^{\mathbb K}(t), z_r^{\mathbb F}(t)|\chi_{m}(t) = r, \bm Z^{\mathbb K,t-1}, {x}_m^{\mathbb F}(t) \right) \nonumber \\
&=& C_{m,r}^{\mathbb K}(t) C_{m,r}^{\mathbb F}(t),
\end{eqnarray}
where $C_{m,r}^{\mathbb K}(t)$ and $C_{m,r}^{\mathbb F}(t)$ are computed by \eqref{eq:likelihoodK} and \eqref{eq:likelihoodF}, respectively. Similar to the original NN-JPDAF \cite{muvsicki2007efficient}, the association probability of track $m$ and measurement $r$ at time $t$ can be calculated by 
\begin{eqnarray}
\label{eq:beta2}
\beta_{m,r}^{\mathbb K,\mathbb F}(t) &=&  P(\chi_{m}(t)=r | \bm Z^{\mathbb K,t}, {x}_m^{\mathbb F}(t) ) \nonumber \\
&\approx& \frac{C_{m,r}^{\mathbb K,\mathbb F}(t)}{D_m^{\mathbb K,\mathbb F}(t)+E_r^{\mathbb K,\mathbb F}(t)-C_{m,r}^{\mathbb K,\mathbb F}(t)+B},
\end{eqnarray}
where $D_m^{\mathbb K,\mathbb F}(t)$ and $E_r^{\mathbb K,\mathbb F}(t)$ are computed by \eqref{eq:D} and \eqref{eq:E} with $C_{m,r}^{\mathbb K}(t)$ replaced by $C_{m,r}^{\mathbb K,\mathbb F}(t)$.

Now we focus on estimating the feature signals which are required for the refinement of the association probability. As is analyzed in the previous section, the assignment between the targets and the feature measurements is missing, which makes it hard to estimate the feature signal and frequency. Note that if we apply the NN-JPDAF algorithm with only kinematic measurements, the algorithm can provide a rough assignment of the measurements to the targets, i.e., the estimate $\hat \chi_m(t)$ for $t=0,1,...,N-1$. For the convenience of derivation, we denote $\tilde z_m^{\mathbb F}(t) = z_{\hat \chi_m(t)}^{\mathbb F}(t)$ for $\hat \chi_m(t) \ne 0$ and $e_m^{\mathbb F}(t) = \tilde z_m^{\mathbb F}(t) - x_m^{\mathbb F}(t) - w_m^{\mathbb F}(t)$ as the association error. If the association is correct, then $e_m^{\mathbb F}(t) = 0$, otherwise $e_m^{\mathbb F}(t) \neq 0$.

Besides mis-associations, missed detections also exist in tracking, and sometimes there is no measurement associated to the $m$-th target. In general, we denote $\Omega_m = \left\{ t_1, ... ,t_{\alpha_m} \right\}$ as the set of time steps that $\hat \chi_m(t) \ne 0$ where $\alpha_m$ is the number of samples. The associated feature measurement vector, measurement noise vector and association error vector are defined as $\bm{\tilde z}_m^{\mathbb F} = \left[ \tilde z_m^{\mathbb F}(t_1), \tilde z_m^{\mathbb F}(t_2), ..., \tilde z_m^{\mathbb F}(t_{\alpha_m}) \right]^T$, $\bm w_m^{\mathbb{F}} = \left[ w_m^{\mathbb F}(t_1), w_m^{\mathbb F}(t_2), ..., w_m^{\mathbb F}(t_{\alpha_m}) \right]^T$ and  $\bm e_m^{\mathbb F} = \left[ e_m^{\mathbb F}(t_1), e_m^{\mathbb F}(t_2), ..., e_m^{\mathbb F}(t_{\alpha_m}) \right]^T$, respectively. Then we have
\begin{eqnarray}
\bm{\tilde z}^{\mathbb F}_m=\bm x_m^{\mathbb F}(\Omega_m)+\bm e_m^{\mathbb F}+\bm w_m^{\mathbb{F}},
\end{eqnarray}
where $\bm x_m^{\mathbb F}(\Omega_m)$ denotes the subvector of $\bm x_m^{\mathbb F}$ composed of elements with indices $\Omega_m$. To extract the feature of the target, we need to reconstruct the feature signal $\bm x_m^{\mathbb F}$ and estimate its spectrum $(\bm c_m, \bm f_m, \bm \phi_m)$ based on the measurements $\bm{\tilde z}^{\mathbb F}_m$. Obviously, the missed detections, mis-associations and the sensing noise exist in the measurements, which makes the problem quite challenging. We will exploit the following two types of sparsity to solve the problem:
\begin{enumerate}
	\item{The signal $\bm x_m^{\mathbb F}$ consists of only a few frequencies, and the number of frequencies is much smaller than the number of samples, i.e., $K_m \ll \alpha_m$.}
	\item{If we can maintain the track successfully, the number of mis-associations should be small compared with the number of time steps $N$.}
\end{enumerate}

Define the number of mis-associations for the $m$-th target as $s_m$, then we have $\| \bm e_m^{\mathbb F} \|_0 = s_m$. Although the $\ell_0$-norm constraint enhances the sparsity of the solution, it usually results in an NP-hard optimization problem. Therefore, we use the $\ell_1$-norm regularization instead, i.e., $\| \bm e_m^{\mathbb F} \|_1 = \sum\limits_{l = 1}^{{\alpha _m}} {|e_m^{\mathbb F}({t_l})|}$. 

The feature signal is a linear combination of complex sinusoids with arbitrary phases, where the frequencies do not fall onto discrete grids. Therefore, it cannot be directly formulated by using the $\ell_1$-norm. A regularizer that encourages a sparse combination of such sinusoids is the atomic norm \cite{chandrasekaran2012convex}. Define an atom $\bm a(f,\phi) \in \mathbb C^N$ as
\begin{eqnarray}
\label{eq:atom}
\bm a(f,\phi) = [e^{i\phi}, e^{i(2\pi f + \phi)}, ..., e^{i(2\pi f (N-1) + \phi)}]^T,
\end{eqnarray}
where $f\in [0,1]$ and $\phi \in [0, 2 \pi)$. Then the feature signal $\bm x_m^{\mathbb F}$ can be written as
\begin{eqnarray}
\bm x_m^{\mathbb F} = \sum\limits_{k = 1}^{K_m} {{c_m(k)} \bm a(f_m(k),\phi_m(k))}.
\end{eqnarray}
Denote the set of atoms as ${\cal A}= \left\{ \bm a (f, \phi): f \in [0,1], \phi \in [0,2\pi) \right\}$.
\begin{definition}\cite{tang2013compressed}
	The atomic norm $\| \cdot \|_{\cal A}$ is defined by:
	\begin{eqnarray}
	\label{eq:atomicnorm}
	\| \bm x \|_{\cal A} &=& \inf \left\{ r>0: \bm x \in r \cdot \text{conv}({\cal A}) \right\}, \nonumber \\
	&=&  \inf \limits_{\scriptstyle{c_k} \ge 0\hfill\atop
		{\scriptstyle{\phi(k)} \in [0,2\pi )\hfill\atop
			\scriptstyle{f(k)} \in [0,1)\hfill}} \left\{ {\sum\limits_k {{c_k}} : \bm x = \sum\limits_k {{c_k}} \bm a({f(k)},{\phi(k)})} \right\}.
	\end{eqnarray}
\end{definition}
The atomic norm can enforce sparsity in the atom set $\cal A$ \cite{tang2013compressed,chandrasekaran2012convex}. On this basis, we can formulate the following optimization problem\footnote{The optimization can also be interpreted as maximum a posteriori probability estimate with sparseness-promoting prior $p \left( \bm x_m^\mathbb{F}|\gamma_m \right) = \delta_{\gamma_m} \exp \left( -(2\gamma_m/\sigma_m^2) \| \bm x_m^\mathbb{F} \|_{\cal A}\right)$ and $p \left( \bm e_m^\mathbb{F}|\lambda_m \right) = \delta_{\lambda_m} \exp \left( -(2\lambda_m/\sigma_m^2) \| \bm e_m^\mathbb{F} \|_1 \right)$. Here $\delta_{\gamma_m}$ and $\delta_{\lambda_m}$ are the normalized factors.} for estimating the feature signal and the association error for the $m$-th target.
\begin{eqnarray}
\label{eq:obj}
\min_{\bm x_m^{\mathbb F}, \bm e_m^{\mathbb F}} \gamma_m \|\bm x_m^{\mathbb F} \|_{\cal A}+\lambda_m \|\bm e_m^{\mathbb F}\|_1 + \frac{1}{2} \|\bm{\tilde z}_m^{\mathbb F}-\bm x_m^{\mathbb F}(\Omega_m)-\bm e_m^{\mathbb F}\|_2^2,
\end{eqnarray}
where $\gamma_m>0$ and $\lambda_m>0$ are weight factors. In practice, we set $\gamma_m \simeq \sigma_m \sqrt{N \log{N}}$ and $\lambda_m \simeq \frac{\gamma_m}{\sqrt{\alpha_m}}$. The way of solving \eqref{eq:obj} will be described in the next section.

With the estimated feature signal, the association probability of the measurements can be computed by using both kinematic and feature information. Note that the original NN-JPDAF is a recursive algorithm for MTT \cite{pulford2005taxonomy}. The kinematic state at time $t$ is estimated based on the measurements at time $t$ and the previous state estimate at time $t-1$. However, the feature estimation at time $t$ is based on the feature measurements over time steps $0,1,..., N-1$. To incorporate the feature information of the targets, we extend the NN-JPDAF to a batch processing tracking algorithm. 

Similar to the other batch tracking algorithms such as the probabilistic multi-hypothesis tracker (PMHT) \cite{willett2002pmht,crouse2009critical}, there can be overlap between adjacent batches. The kinematic state and covariance of the first time step in a batch are initialized by the corresponding state and covariance in the previous batch. The kinematic states in other time steps within the batch are estimated through filtering. The feature in the overlapped area are initialized by the estimates of the previous batch. In the overlapped area, the data association uses both the kinematic and feature measurements while the data association only uses the kinematic measurements outside the overlapped area.

\begin{figure}[htbp]
	\centering
	\subfloat{\includegraphics[width=4in]{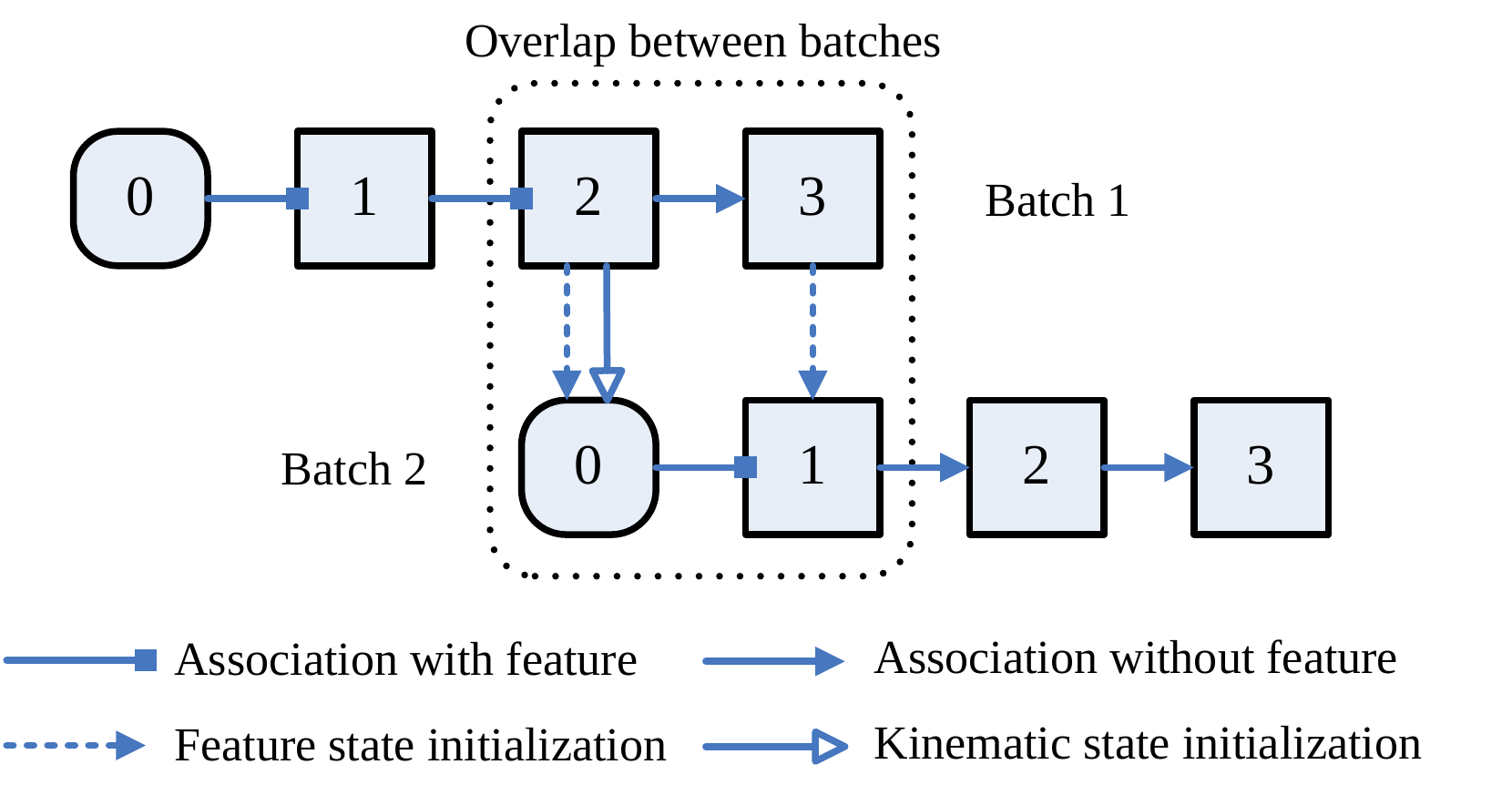}}
	\caption{Example for batch sliding and state initialization.}
	\label{fig:batch}
\end{figure}

Fig. \ref{fig:batch} gives an example for the batch sliding and state initialization. As can be seen from the figure, Batch 1 and Batch 2 both contain four time steps and the overlapped area contains two time steps. In Batch 1, the algorithm firstly performs the tracking in the entire batch with only the kinematic measurements and obtains a rough estimate of the assignments between measurements and tracks. By solving \eqref{eq:obj}, the algorithm estimates the feature from time 0 to time 3 in Batch 1. Then a re-filtering is made and the data association at time 1 and 2 uses both kinematic and feature measurements. The filtering result at time 2 is used for the initialization of the kinematic state and covariance at time 0 in Batch 2. The feature at time 0, 1 in Batch 2 are initialized by the feature at time 2, 3 of the Batch 1. In Batch 2, the data association at time 1 uses both kinematic and feature information, while the data association at time 2, 3 uses only the kinematic information.

As there is overlap between adjacent batches, the feature signal estimated in the previous batch can be used as the prior information for the feature signal in the current batch. Suppose the overlapped area is $A$ time steps in length, then we define $\Xi=[0,1,...,A-1]$ as the overlapped time steps, and $\bm{\bar x}_m^{\mathbb F} = [\bar x_m^{\mathbb F}(0),\bar x_m^{\mathbb F}(1),...,\bar x_m^{\mathbb F}(A-1)]^T$ as the feature estimate given by the previous batch. These previous estimates serve as an additional constraint to the optimization problem in \eqref{eq:obj} for feature estimation in the current batch, i.e.,
\begin{eqnarray}
	\label{eq:obj2}
	\min_{\bm x_m^{\mathbb F}, \bm e_m^{\mathbb F}} \gamma_m \|\bm x_m^{\mathbb F} \|_{\cal A}+\lambda_m \|\bm e_m^{\mathbb F}\|_1 + \frac{1}{2}\|\bm{\tilde z}_m^{\mathbb F}-\bm x_m^{\mathbb F}(\Omega_m)-\bm e_m^{\mathbb F}\|_2^2 + \frac{\zeta_m}{2}\|\bm x_m^{\mathbb F}(\Xi)-\bm {\bar x}_m^{\mathbb F}\|_2^2,
\end{eqnarray}
for the $m$-th target. $\zeta_m$ is the weight of the additional $\ell_2$-penalty. We denote the solutions to \eqref{eq:obj} and \eqref{eq:obj2} as $\bm{\hat x}_m^{\mathbb F} = [\hat x_m^{\mathbb F}(0),\hat x_m^{\mathbb F}(1),...,{\hat x}_m^{\mathbb F}(N-1)]^T$ and $\bm{\hat e}_m^{\mathbb F}=[\hat e_m^{\mathbb F}(0),\hat e_m^{\mathbb F}(1),...,\hat e_m^{\mathbb F}(\alpha_m-1)]^T$.

With the estimate of feature signal, the association probability can be calculated according to \eqref{eq:likelihoodF}, \eqref{eq:likelihoodJ} and \eqref{eq:beta2} with $x_m^\mathbb{F}(t)$, $\sigma_m^2$ replaced by $\hat x_m^\mathbb{F}(t)$ and $\tilde \sigma_m^2(t)$, respectively. Here $\tilde \sigma_m^2(t)$ is the variance of $x_m^{\mathbb F}(t)+w_m^{\mathbb F}(t)-\hat x_m^{\mathbb F}(t)$. The rest of the association is the same as that of the original NN-JPDAF: the algorithm iterates until the termination condition is satisfied, and the measurement-to-track pair that gives the largest $\beta_{m,r}^{\mathbb K,\mathbb F}(t)$ is associated in each iteration. The tracks are updated according to \eqref{eq:xupda}, \eqref{eq:Pupda}, \eqref{eq:innovation} and \eqref{eq:gain}.

\section{Sparsity-based Feature Extraction Algorithm}
\subsection{Feature Estimation and Frequency Estimation}
\label{sec:feaest}

In the previous section, a sparsity-based optimization formulation is proposed to estimate the feature signal of the targets. In this section, we develop the way to solve the optimization problem and estimate the frequency of the feature. We will use an equivalent form of the atomic norm for the atom set $\cal A$ \cite{tang2013compressed}:
\begin{eqnarray}
\label{eq:atomic}
\left\| \bm x \right\|_{\cal A} = \mathop {\inf }\limits_{\bm u,\theta } \left\{ \begin{array}{l}
\frac{1}{{2N}}{\rm{Tr}}({\rm Toep}(\bm u)) + \frac{\theta }{2},\\
{\rm s.t.} \left[ {\begin{array}{*{20}{c}}
	{{\rm Toep}(\bm u)}& \bm x\\
	{{\bm x^H}}&\theta 
	\end{array}} \right] \succeq 0
\end{array} \right\},
\end{eqnarray}
where $\rm{Tr}(\cdot)$ is the trace of the input matrix, $\text{Toep}(\cdot)$ denote the Toeplitz matrix whose first column is the input vector. Applying \eqref{eq:atomic}, we transform \eqref{eq:obj} to the following semidefinite program (SDP):
\begin{eqnarray}
\label{eq:sdp}
\min \limits_{\bm u_m,\bm x_m^{\mathbb F}, \bm e_m^{\mathbb F},\theta_m } && \frac{\gamma_m }{2} \left( u_m(1)+\theta_m\right) +\lambda_m \|\bm e_m^{\mathbb F}\|_1 + \frac{1}{2} \|\bm{\tilde z}_m^{\mathbb F}-\bm x_m^{\mathbb F}(\Omega_m)-\bm e_m^{\mathbb F}\|_2^2,\\
\text{s.t.}&&\left[ {\begin{array}{*{20}{c}}
	{\text{Toep}(\bm u_m)}& \bm x_m^{\mathbb F}\\
	{(\bm x_m^{\mathbb F})^H}&\theta_m 
	\end{array}} \right] \succeq 0. \nonumber
\end{eqnarray}
The above problem is convex, so can be solved efficiently using a convex solver. Obviously, the solution $\bm{\hat x}_m^{\mathbb F}$ is the estimate of feature signal, and the mis-associations can be estimated by locating the non-zero elements in $\bm{\hat e}_m^{\mathbb F}$. 

Solving \eqref{eq:sdp} does not directly provide the estimates of the frequencies. In many cases, the frequency of the feature signal is of great interest, especifically for target identification. The frequency of the signal can be estimated by solving the dual problem of \eqref{eq:obj}. The dual norm of $\| \cdot \|_{\cal A}$ is defined as \cite{tang2013compressed}
\begin{eqnarray}
	\| \bm q \|_{\cal A}^* = \sup_{\| \bm x \|_{\cal A} \leq 1} \langle \bm q , \bm x \rangle_{\mathbb R}.
\end{eqnarray}
Following a standard Lagrangian analysis \cite{chandrasekaran2012convex}, the dual problem of \eqref{eq:obj} is given by
\begin{eqnarray}
\label{eq:dual}
\max_{\bm q_m} &&{\left\langle {{\bm q_m}({\Omega _m}),\bm{\tilde z}_m^{\mathbb F}} \right\rangle _{\mathbb R}} - \frac{1}{2}\left\| {\bm q_m} \right\|_2^2, \\
\text{s.t.} && \|\bm q_m \|_{\cal A}^* \leq \gamma_m \nonumber,\\
&& \|\bm q_m \|_\infty \leq \lambda_m,\nonumber\\
&& \bm q_m(\Omega_m^c)=0, \nonumber
\end{eqnarray}
where $\bm q_m  \in {\mathbb C}^{N}$ is the dual variable. According to Theorem 4.24 in \cite{dumitrescu2007positive}, $\|\bm q_m \|_{\cal A}^* \leq \gamma_m$ is equivalent to the following constraints
\begin{eqnarray}
\label{eq:equaDual}
&& \left[ {\begin{array}{*{20}{c}}
	\bm U_m& \bm q_m\\
	{{\bm q_m^H}}&1
	\end{array}} \right] \succeq 0, \nonumber \\
&& \bm U_m \in {\mathbb C}^{N \times N} \text{ is a Hermitian matrix}, \nonumber \\
&& \text{Tr}(\bm U_m)=\gamma_m^2, \text{Tr}_j (\bm U_m)=0, j = 1,2,...,N-1, \nonumber
\end{eqnarray}
where $\text{Tr}_j(\bm U)$ denotes the sum of the elements on the $j$-th subdiagonal of $\bm U$. Hence, the dual problem can also be transformed to an SDP. In practice, solving the dual problem is equivalent to solving the primal problem, and most solvers can directly return a dual optimal solution when solving the primal problem.

Let $\bm{\hat q}_m=[\hat q_m(0),\hat q_m(1),...,\hat q_m(N-1)]^T$ be the solution to \eqref{eq:dual}. The following lemma can be used to identify the frequencies of the primal solution from a dual solution.
\begin{lemma}
Suppose $\bm{\hat x}_m^{\mathbb F} = \sum \limits_{k = 1}^{{K_m}} {{\tilde c_m}(k) \bm a({\tilde f_m(k)},\tilde \phi_m(k))}$ and $\bm{\hat e}_m^{\mathbb F}$ are the primal solutions to \eqref{eq:obj}, then the dual polynomial $Y_m(f)=\langle \bm{\hat q}_m, \bm a(f,0) \rangle$ satisfies
\begin{eqnarray}
\label{eq:dualc}
&& Y_m(\tilde f_m(k)) = \gamma_m e^{i \tilde \phi_m(k)}, k=1,2,...,K_m,\\
\label{eq:dualc2}
&& \hat q_m(\Omega_m(j)) = \lambda_m \text{sign}(\hat e_m^{\mathbb F}(\Omega_m(j))), \forall \hat e_m^{\mathbb F}(j) \neq 0, j=1,2,...,\alpha_m.
\end{eqnarray}
where $\text{sign}(e)=\frac{e}{|e|}$.
\end{lemma}

\vspace{.2cm}

The proof of Lemma 1 is given in the appendix. According to \eqref{eq:dualc}, the recovered frequencies of the feature signal can be obtained by identifying points where the dual polynomial has modulus $\gamma_m$. Moreover, the dual solution provides another way to detect the mis-associations: in places with mis-associations, the amplitude of the dual solution equals to $\lambda_m$.

\begin{figure}[htbp]
	\centering
	\subfloat[][]{\includegraphics[width=2.3in]{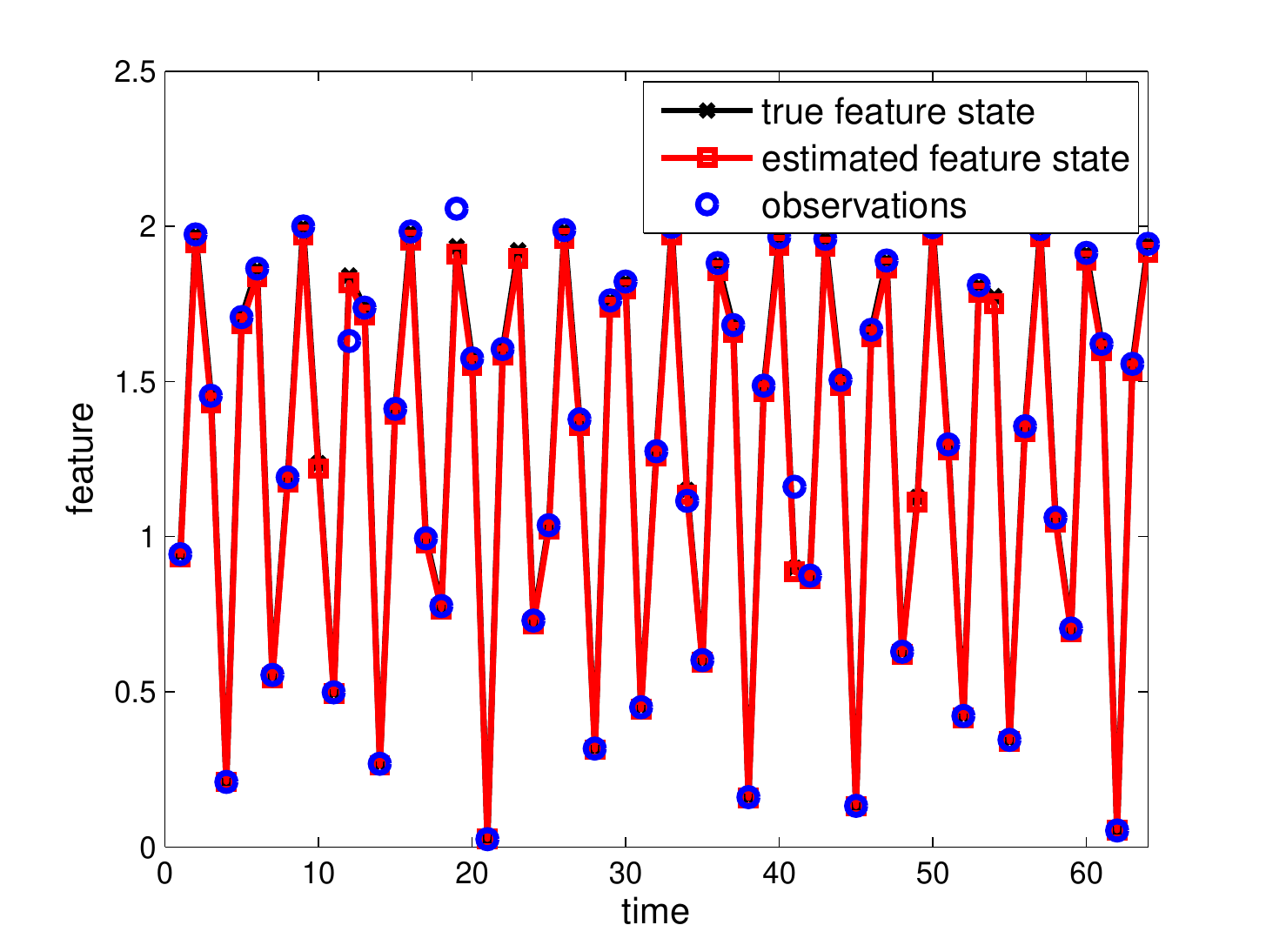}}
	\subfloat[][]{\includegraphics[width=2.3in]{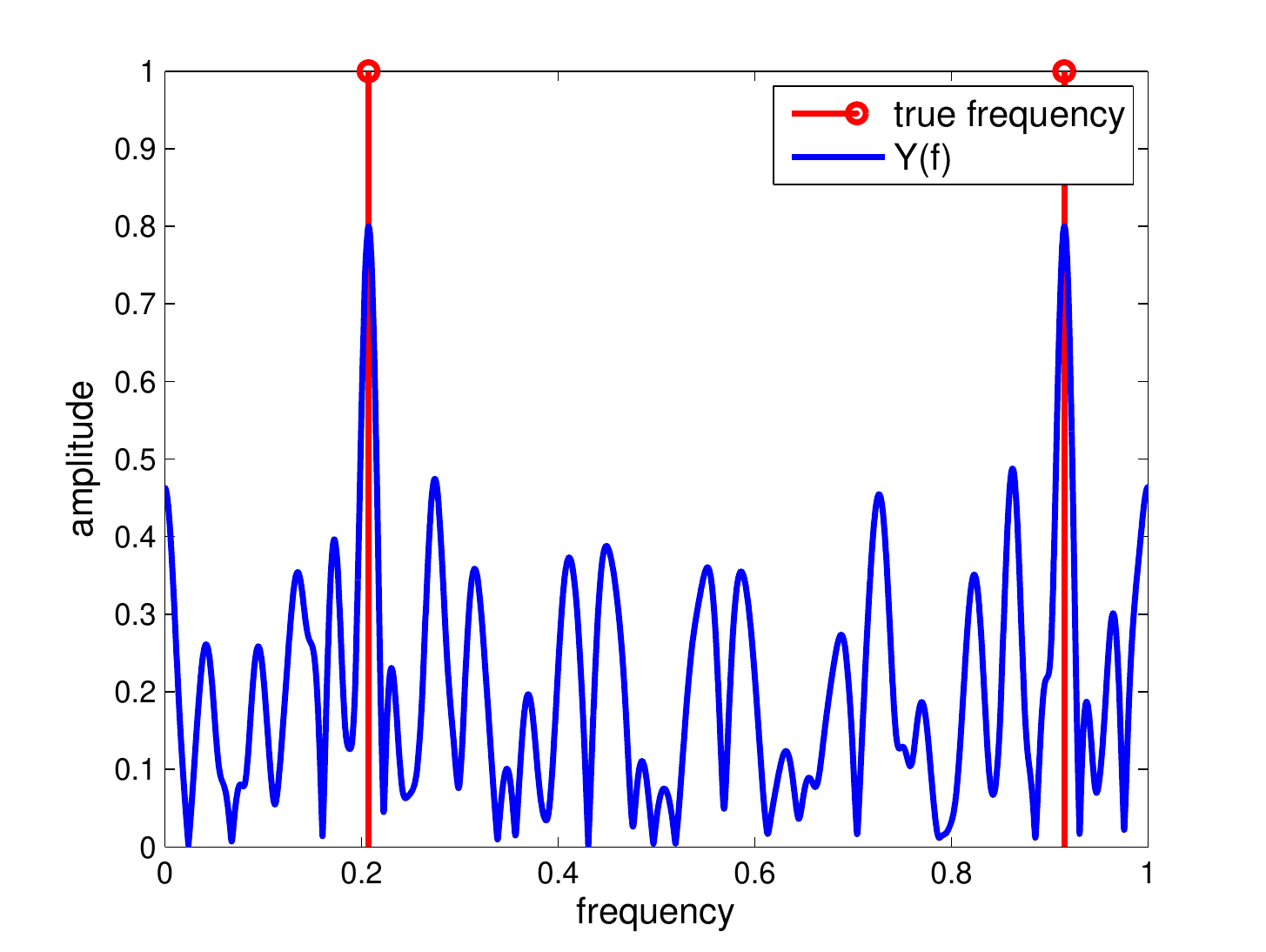}}
	\subfloat[][]{\includegraphics[width=2.3in]{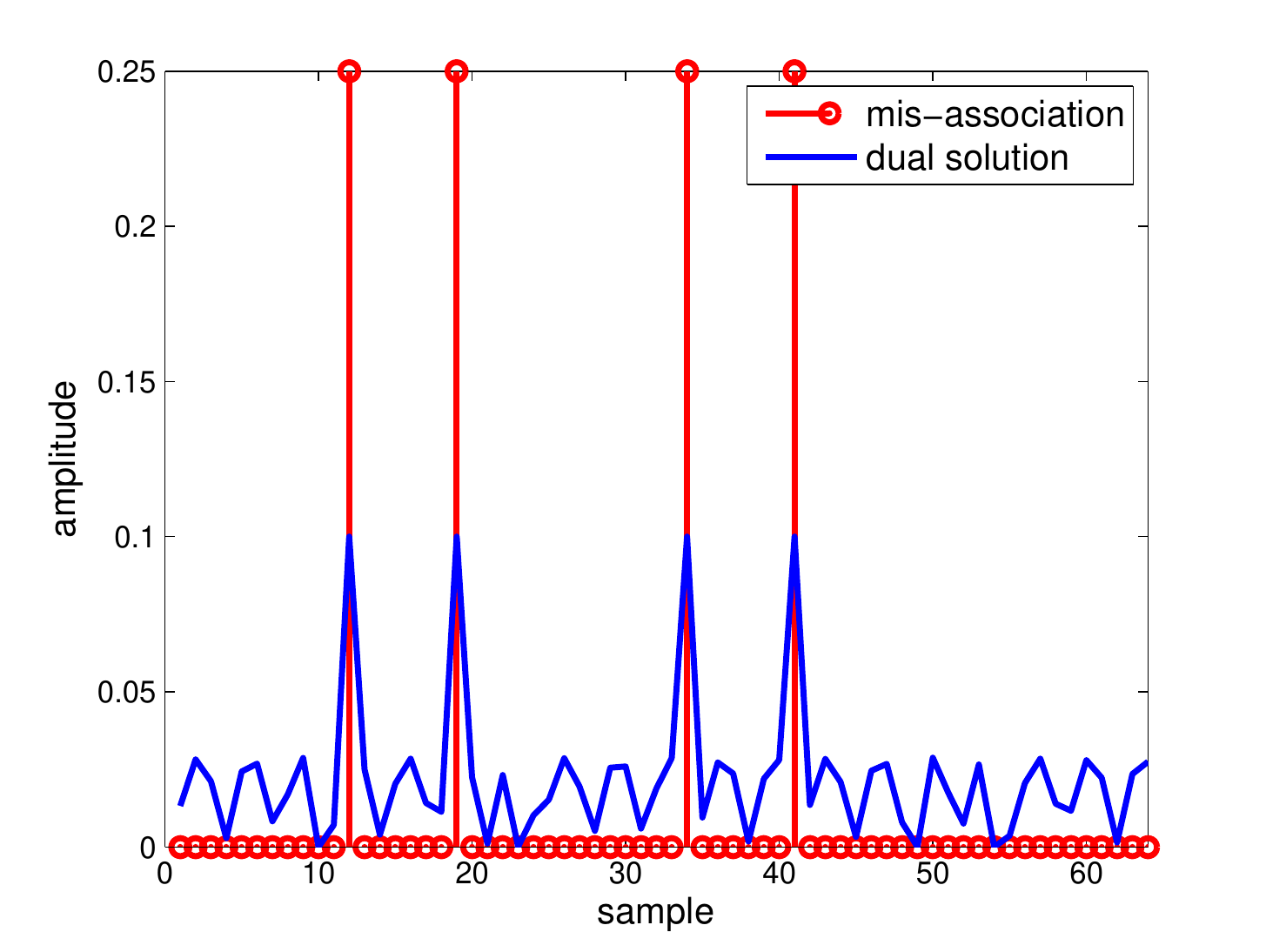}}
	\caption{(a) Plots of feature signal, feature measurement, and the estimated signal based on solving \eqref{eq:sdp}. (b) Plots of dual polynomial and the frequency of the signal. (c) Plots of dual solution and the mis-association.}
	\label{fig:dual}
\end{figure}

Fig. \ref{fig:dual} gives an example of the feature extraction and mis-association detection through the dual solution. The feature of the target has two spikes in frequency domain with equal amplitude. For simplicity, we assume there is no noise in this scenario and $\gamma_m$ in \eqref{eq:obj} is set to 0.8. We track a target over 64 frames, while 4 frames of measurements are missed. Hence,  $\alpha_m=60$ and $\lambda_m$ is set as 0.1 in our simulation. Meanwhile, there are 4 mis-associations. For target classification, we usually need to estimate the frequency of the feature with a few number of measurements corrupted by mis-association. As can be seen from Fig. \ref{fig:dual}b and Fig. \ref{fig:dual}c, the recovered frequencies are obtained by identifying points where the dual polynomial has magnitude $\gamma_m$, and the position of mis-association can be detected by identifying the points where the dual solution has magnitude $\lambda_m$. 

For the feature estimation with overlap between the batches, the formulation given in \eqref{eq:obj2} is used for feature estimation, which also has an equivalent SDP form. Specifically, it can be reformulated by \eqref{eq:sdp} with an extra term $\frac{\zeta_m}{2}\|\bm x_m^{\mathbb F}(\Xi)-\bm {\bar x}_m^{\mathbb F}\|_2^2$ in the objective function. We introduce the dual problem for \eqref{eq:obj2} to estimate the frequency of the feature signal. Except for $\bm q_m$, we need to introduce a new dual variable $\bm g_m=[g_m(1),g_m(2),...,g_m(N-1)]^T$. Then the dual problem for \eqref{eq:obj2} becomes
\begin{eqnarray}
\label{eq:dual2}
\max_{\bm q_m, \bm g_m} &&\langle \bm q_m(\Omega_m) , \bm{\tilde z}_m^{\mathbb F} \rangle_{\mathbb R} + \langle \bm g_m(\Xi) , \bm {\bar x}_m^{\mathbb F} \rangle_{\mathbb R} - \frac{1}{2} \|\bm q_m\|_2^2 - \frac{1}{2 \zeta_m} \|\bm g_m\|_2^2, \\
\text{s.t.} && \|\bm q_m \|_{\cal A}^* \leq \gamma_m, \|\bm g_m \|_{\cal A}^* \leq \gamma_m \nonumber,\\
&& \|\bm q_m \|_\infty \leq \lambda_m, \nonumber\\
&& \bm q_m(\Omega_m^c)=0, \bm g_m(\Xi^c)=0. \nonumber
\end{eqnarray}
Let $(\bm {\hat x}_m^{\mathbb F},\bm {\hat e}_m^{\mathbb F})$ and $(\bm{\hat q}_m, \bm{\hat g}_m)$ be a primal-dual pair of solutions to \eqref{eq:obj2} and \eqref{eq:dual2}. Following the similar proof of Lemma 1, it can be shown that the dual solution satisfies \eqref{eq:dualc} and \eqref{eq:dualc2}, so the feature frequency can be estimated by checking the dual polynomial of the problem.

\subsection{Accelerated Feature Extraction}
\label{sec:admm}
According to the previous section, our objective function \eqref{eq:obj} is equivalent to the SDP given by \eqref{eq:sdp} which can be solved by off-the-shelf solvers such as SeDuMi \cite{sturm1999using} and SDPT3 \cite{toh1999sdpt3}. However, these solvers tend to be slow, especially for large problems. To meet the requirement of real time signal processing, we provide a reasonably fast method for solving this SDP via the alternating direction method of multipliers (ADMM). To put our problem in an appropriate form for ADMM, rewrite \eqref{eq:sdp} as
\begin{eqnarray}
	\label{eq:sdp2}
	\min \limits_{\bm u_m,\bm x_m^{\mathbb F}, \bm e_m^{\mathbb F},\theta_m } && \frac{\gamma_m }{2} \left( u_m(1)+\theta_m\right) +\lambda_m \|\bm e_m^{\mathbb F}\|_1 + \frac{1}{2} \|\bm{\tilde z}_m^{\mathbb F}-\bm x_m^{\mathbb F}(\Omega_m)-\bm e_m^{\mathbb F}\|_2^2,\\
	\text{s.t.}&& \bm \Psi_m = \left[ {\begin{array}{*{20}{c}}
			{\text{Toep}(\bm u_m)}& \bm x_m^{\mathbb F}\\
			{(\bm x_m^{\mathbb F})^H}&\theta_m 
		\end{array}} \right], \nonumber \\
		&& \bm \Psi_m \succeq 0  \nonumber,
\end{eqnarray}
and dualize the equality constraint via an Augmented Lagrangian:
\begin{eqnarray}
	\label{eq:AL}
	{\cal L}_\rho(\theta_m,\bm u_m,\bm x_m^{\mathbb F},\bm e_m^{\mathbb F},\bm \Upsilon_m,\bm \Psi_m) &=&\frac{\gamma_m }{2} \left( u_m(1)+\theta_m\right) +\lambda_m \|\bm e_m^{\mathbb F}\|_1 + \frac{1}{2} \|\bm{\tilde z}_m^{\mathbb F}-\bm x_m^{\mathbb F}(\Omega_m)-\bm e_m^{\mathbb F}\|_2^2 \nonumber \\
	&& + \left\langle {{\bm \Upsilon _m},\bm \Psi_m  - \left[ {\begin{array}{*{20}{c}}
	{\text{Toep}(\bm u_m)}& \bm x_m^{\mathbb F}\\
	(\bm x_m^{\mathbb F})^H&{{\theta _m}}
	\end{array}} \right]} \right\rangle \nonumber \\
	&& + \frac{\rho}{2} \left \| {\bm \Psi_m  - \left[ {\begin{array}{*{20}{c}}
	{\text{Toep}(\bm u_m)}& \bm x_m^{\mathbb F}\\
	(\bm x_m^{\mathbb F})^H&{{\theta _m}}
	\end{array}} \right] }\right\|_F^2.
	\end{eqnarray}
ADMM then consists of the update steps \cite{boyd2011distributed}:
\begin{eqnarray}
\label{eq:UPDA}
(\theta_m^{l+1},\bm x_m^{\mathbb F,l+1},\bm e_m^{\mathbb F,l+1},\bm u_m^{l+1}) &=& \arg\min_{\theta_m,\bm x_m^{\mathbb F},\bm e_m^{\mathbb F},\bm u_m}{\cal L}_{\rho}(\theta_m,\bm u_m,\bm x_m^{\mathbb F},\bm e_m^{\mathbb F},\bm \Upsilon_m^l,\bm \Psi_m^l),\\
\label{eq:UPDA2}
\bm \Psi_m^{l+1} &=& \arg\min_{\bm \Psi_m}{\cal L}_{\rho}(\theta_m^{l+1},\bm u_m^{l+1},\bm x_m^{\mathbb F,l+1},\bm e_m^{\mathbb F,l+1},\bm \Upsilon_m^l,\bm \Psi_m),\\
\label{eq:UPDAUpsilon}
\bm \Upsilon_m^{l+1} &=& \bm \Upsilon_m^l + \rho \left( {\bm \Psi_m^{l+1}  - \left[ {\begin{array}{*{20}{c}}
{\text{Toep}(\bm u_m^{l+1})}& \bm x_m^{\mathbb F,l+1}\\
(\bm x_m^{\mathbb F,l+1})^H&{{\theta _m^{l+1}}}
\end{array}} \right] } \right).
\end{eqnarray}
			
Now we explain the update \eqref{eq:UPDA} and \eqref{eq:UPDA2} in detail. For the convenience of our description, the following partitions are introduced:
\begin{eqnarray}
\bm \Psi _m^l = \left[ {\begin{array}{*{20}{c}}
		{\bm \Psi _{m,0}^l}&{\bm \psi _{m,1}^l}\\
		{(\bm \psi _{m,1}^l)^H}&{\Psi _{m,N+1}^l}
		\end{array}} \right],
\end{eqnarray}
\begin{eqnarray}
\bm \Upsilon _m^l = \left[ {\begin{array}{*{20}{c}}
		{\bm \Upsilon _{m,0}^l}&{\bm \upsilon _{m,1}^l}\\
		{(\bm \upsilon _{m,1}^l)^H}&{\Upsilon _{m,N+1}^l}
	\end{array}} \right],
\end{eqnarray}
where $\bm \Psi _{m,0}^l$ and $\bm \Upsilon _{m,0}^l$ are $N \times N$ matrices, $\bm \psi _{m,1}^l$ and $\bm \upsilon _{m,1}^l$ are $N \times 1$ vectors, $\Psi _{m,N+1}^l$ and $\Upsilon _{m,N+1}^l$ are scalars. Computing the derivative of ${\cal L}_{\rho}(\theta_m,\bm u_m,\bm x_m^{\mathbb F},\bm e_m^{\mathbb F},\bm \Upsilon_m^l,\bm \Psi_m^l)$ with respect to $\bm x_m^{\mathbb F}$, $\theta_m$ and $\bm u_m$, we have
\begin{eqnarray}
\nabla_{\bm x_m^{\mathbb F}(\Omega_m)} {\cal L}_{\rho} &=& \bm x_m^{\mathbb F}(\Omega_m) + \bm e_m^{\mathbb F} - \bm z_m^{\mathbb F} - 2 \bm \upsilon_{m,1}^l(\Omega_m) + 2\rho (\bm x_m^{\mathbb F}(\Omega_m) - \bm \psi_{m,1}^l(\Omega_m)),\\
\nabla_{\bm x_m^{\mathbb F}(\Omega_m^c)} {\cal L}_{\rho} &=& - 2 \bm \upsilon_m^l(\Omega_m^c) + 2\rho (\bm x_m^{\mathbb F}(\Omega_m^c) - \bm \psi_{m,1}^l(\Omega_m^c)),\\
\nabla_{\theta _m} {\cal L}_{\rho} &=& \frac{{{\gamma _m}}}{2} - \Upsilon _{m,N + 1}^l + \rho ({\theta _m} - {\Psi _{m,N + 1}^l}),\\
\nabla_{{u_m}(j)}{\cal L}_{\rho} &=& \left\{ \begin{array}{l}
\frac{{{\gamma _m}}}{2} + N\rho {u_m}(j) - {\rm{Tr}}(\rho \bm \Psi _{m,0}^l + \bm \Upsilon _{m,0}^l),j = 1,\\
2(N - j + 1)\rho {u_m}(j) - 2{\rm{Tr}}_{j-1}(\rho \bm \Psi _{m,0}^l + \bm \Upsilon _{m,0}^l),j = 2,3,...,N,
\end{array} \right.
\end{eqnarray}

By setting the derivatives to 0, $\bm x_m^{\mathbb F,l+1}$, $\theta_m^{l+1}$ and $\bm u_m^{l+1}$ can be updated by:
\begin{eqnarray}
\label{eq:UPDAx1}
\bm x_m^{\mathbb F,l+1}(\Omega_m) &=& \frac{1}{2\rho+1} (\bm{\tilde z}_m^{\mathbb F} - \bm e_m^{\mathbb F,l} + 2\rho \bm \psi_{m,1}^l(\Omega_m) + 2 \bm \upsilon_{m,1}^l(\Omega_m)),\\
\label{eq:UPDAx2}
\bm x_m^{\mathbb F,l+1}(\Omega_m^c) &=&   \bm \psi_{m,1}^l(\Omega_m^c) + \frac{1}{\rho}\bm \upsilon_{m,1}^l(\Omega_m^c),\\
\label{eq:UPDAtheta}
\theta_m^{l+1} &=& \Psi_{m,N+1}^l + \left(\Upsilon_{m,N+1}^l - \gamma_m/2 \right)/\rho,\\
\label{eq:UPDAu}
\bm u_m^{l+1} &=& \bm K \text{Tr}^*(\bm \Psi_{m,0}^l + \bm \Upsilon_{m,0}^l/\rho) - \frac{\gamma_m}{2 N \rho} \bm I_1 ,
\end{eqnarray}
where $\bm I_1 = [1,0,0,...,0]^T$, $\text{Tr}^*$ outputs a vector whose $j$-th element is the trace of the $(j-1)$-th subdiagonal of the input matrix, $\bm K$ is the diagonal matrix with entries
\begin{eqnarray}
K(j,j) =\frac{1}{{N - j + 1}},j = 0,1,...N.
\end{eqnarray}
The term in \eqref{eq:AL} that is related to $\bm e_m^{\mathbb F}$ is $\lambda_m \|\bm e_m^{\mathbb F}\|_1 + \frac{1}{2} \|\bm{\tilde z}_m^{\mathbb F}-\bm x_m^{\mathbb F,l}(\Omega_m)-\bm e_m^{\mathbb F}\|_2^2$. Hence, $\bm e_m^{\mathbb F}$ can be updated by 
\begin{eqnarray}
\bm e_m^{\mathbb F,l+1}=\arg \min_{\bm e_m^{\mathbb F}} \lambda_m \|\bm e_m^{\mathbb F}\|_1 + \frac{1}{2} \|\bm{\tilde z}_m^{\mathbb F}-\bm x_m^{\mathbb F,l}(\Omega_m)-\bm e_m^{\mathbb F}\|_2^2,
\end{eqnarray}
which can be easily achieved by the proximal operator \cite{maleki2013asymptotic}:
\begin{eqnarray}
\bm e_m^{\mathbb F,l+1}=\text{prox}_{\lambda_m}(\bm{\tilde z}_m^{\mathbb F}-\bm x_m^{\mathbb F,l}(\Omega_m)),
\end{eqnarray}
where $\text{prox}_{\lambda}(\bm e) = [\text{prox}_{\lambda}(e_1), \text{prox}_{\lambda}(e_2),...,\text{prox}_{\lambda}(e_N)]^T$ with
\begin{eqnarray}
\label{eq:prox}
\text{prox}_{\lambda}(e_i) = (|e_i|-\lambda) \text{sign}(e_i) \mathbb{I}(|e_i|>\lambda).
\end{eqnarray}	
						
The update of $\bm \Psi _m^l$ is simply the projection onto the positive semidefinite cone
\begin{eqnarray}
\label{eq:UPDAPhi}
\bm \Psi_m^{l+1} = \arg\min_{\bm \Psi_m \succeq 0} \left \| {\bm \Psi_m  - \left[ {\begin{array}{*{20}{c}}
	{\text{Toep}(\bm u_m^{l+1})}& \bm x_m^{\mathbb F,l+1}\\
	(\bm x_m^{\mathbb F,l+1})^H&{{\theta _m^{l+1}}}
    \end{array}} \right] } + \bm \Upsilon_m^{l+1}/\rho \right\|_F^2,
\end{eqnarray}
where projecting a matrix onto the positive definite cone is accomplished by forming an eigenvalue decomposition of the matrix and setting all negative eigenvalues to zero. Noticing that $\bm \Upsilon_m$ corresponds to the dual variables, ADMM also provides the dual solution to \eqref{eq:dual}. Specifically, we have the dual solution $\bm{\hat q}_m = -\frac{\bm{\hat \upsilon}_{m,1}}{2}$.

For the case with overlap between adjacent batches, equation \eqref{eq:obj2} can also be solved fast by using the ADMM. The overflow of the algorithm is the same as that in Section III-D, whereas the update of $\bm x_m^{\mathbb F,l+1}$ is different:
\begin{eqnarray}
\label{eq:UPDAx3}
\bm x_m^{\mathbb F,l+1}(\Omega_m \cap \Xi) &=& \frac{1}{2\rho+\zeta_m+1} (\bm{\tilde z}_m^{\mathbb F}(\Theta_m) - \bm e_m^{\mathbb F,l}(\Theta_m) + \zeta_m \bm{\bar x}_m^{\mathbb F}(\Omega_m \cap \Xi) \nonumber \\
&&+ 2\rho \bm \psi_{m,1}^l(\Omega_m \cap \Xi) + 2 \bm \upsilon_{m,1}^l(\Omega_m \cap \Xi)),\\
\label{eq:UPDAx4}
\bm x_m^{\mathbb F,l+1}(\Omega_m^c \cap \Xi) &=& \frac{1}{2\rho + \zeta_m} (\zeta_m \bm{\bar x}_m^{\mathbb F}(\Omega_m^c \cap \Xi) + 2\rho \bm \psi_{m,1}^l(\Omega_m^c \cap \Xi) + 2 \bm \upsilon_{m,1}^l(\Omega_m^c \cap \Xi)),\\
\label{eq:UPDAx5}
\bm x_m^{\mathbb F,l+1}(\Omega_m \cap \Xi^c) &=& \frac{1}{2\rho+1} (\bm{\tilde z}_m^{\mathbb F}(\Theta_m^c) - \bm e_m^{\mathbb F,l}(\Theta_m^c) + 2\rho \bm \psi_{m,1}^l(\Omega_m \cap \Xi^c) + 2 \bm \upsilon_{m,1}^l(\Omega_m \cap \Xi^c) ),\\
\label{eq:UPDAx6}
\bm x_m^{\mathbb F,l+1}(\Omega_m^c \cap \Xi^c) &=& \rho \bm \psi_{m,1}^l(\Omega_m^c \cap \Xi^c) + \frac{1}{\rho} \bm \upsilon_{m,1}^l(\Omega_m^c \cap \Xi^c),
\end{eqnarray}
where $\Theta_m$ is the set of indices of associated measurements that is inside the overlapped area, $\Theta_m^c$ is the set of indices of associated measurements that is outside the interval of the overlapped area.

\begin{figure}[htbp]
	\centering
	\subfloat{\includegraphics[width=4in]{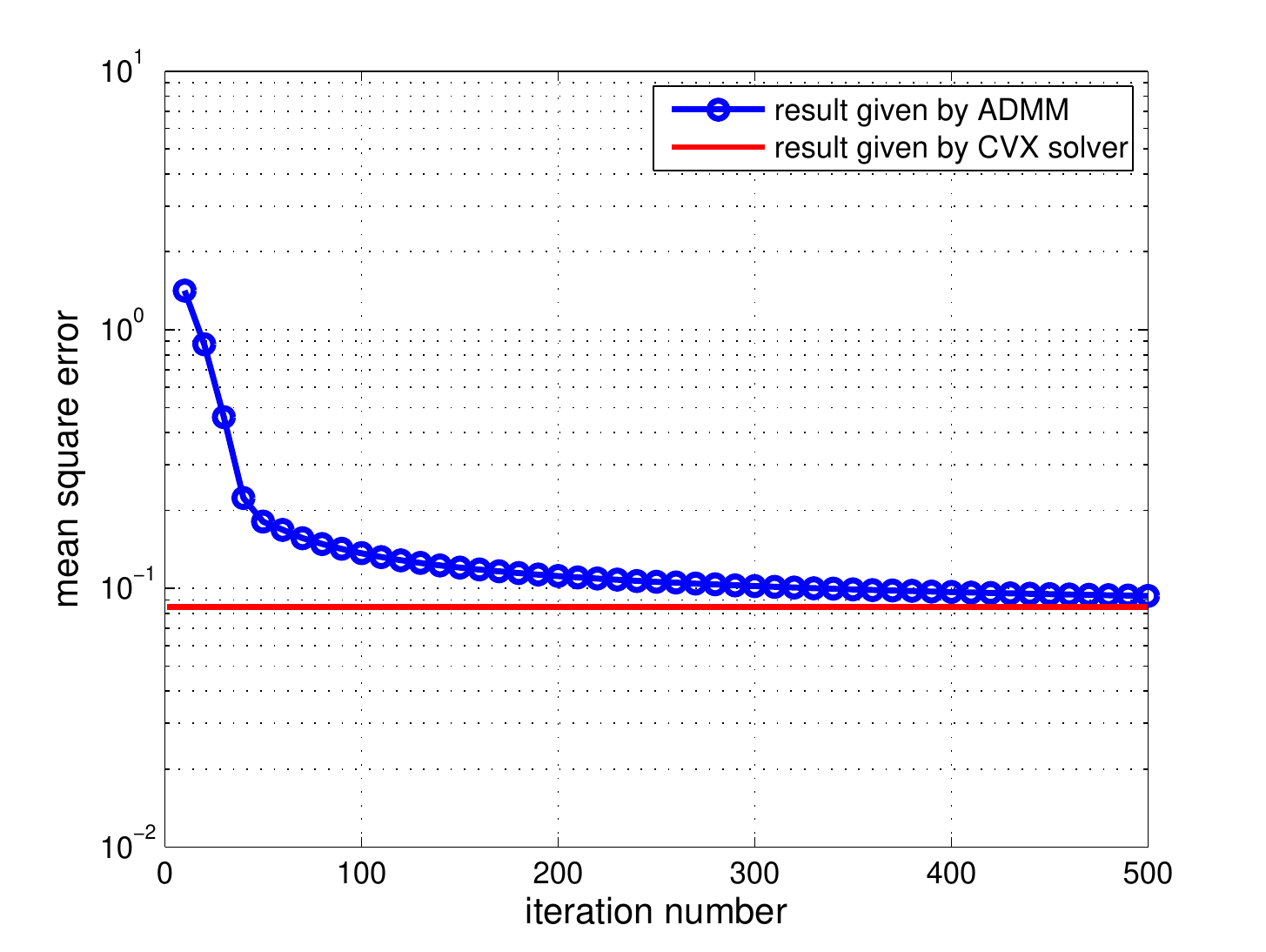}}
	\caption{Convergence behavior of ADMM. The ADMM takes 6.39 seconds with 500 iterations while CVX takes 42.40 seconds.}
	\label{fig:ADMM}
\end{figure}

We illustrate the convergence of the ADMM through a simulation example. For simplicity, we consider a case without overlap between the batches. The dimension of the signal is $N=128$ and the number of measurements is $\alpha_m=100$. The variance of the noise is set as $\sigma_m^2=0.02$. The weighting parameters for the algorithms are set as $\gamma_m=\sigma_m \sqrt{N\log N}$ and $\lambda_m = \frac{\gamma_m}{\sqrt{\alpha_m}}$. We compare the MSE of the proposed ADMM algorithm with that given by directly solving \eqref{eq:sdp} with CVX \cite{cvx}. As can be seen from Fig. \ref{fig:ADMM}, ADMM converges to the solution given by the CVX after 500 iterations. It is worth noting that the ADMM runs much faster than CVX because the calculation in each iteration is in closed-form.

\subsection{Summary of the Feature-aided NN-JPDAF Algorithm}
\label{sec:batch}

For clarity, we summarize the proposed feature-aided NN-JPDAF (FA-NN-JPDAF) algorithm. When there is overlap between the adjacent batches, the pseudo code of the algorithm is given in Algorithm 1. In the pseudo code, $\bm{\bar x}_m^{\mathbb K}(0)$ and $\bm{\bar P}(0)$ are the filtering results given by the previous batch, which are used for initializing the current batch. $\bm {\bar x}_m^{\mathbb F}$ is the estimated feature signal in the overlap area, which can be used to calculate the association weights. The algorithm uses both kinematic and feature measurements for tracking, then the filtering result at time $N-A-1$ is used for initialization of the next batch.

When the tracks are first initialized, there is no previous batch, so the overlap does not exist. In such cases, the kinematic states and covariance can be initialized by track-initiation techniques such as two-point differencing \cite{willett2001integration}. The tracking and feature estimation in the first batch can be implemented according to Algorithm 1, with $A=1$ and $\bm {\hat x}_m^{\mathbb F}$ in line 5 updated by iterating with \eqref{eq:UPDAUpsilon}, \eqref{eq:UPDAx1}-\eqref{eq:UPDAPhi}. 

\begin{algorithm}
	\label{tab:A1}
	\caption{Feature-aided NN-JPDAF algorithm}
	\begin{tabular}{lcl}
		Input $\bm{\bar x}_m^{\mathbb K}(0)$, $\bm{\bar P}_m(0)$, $\bm{\bar x}_m^{\mathbb F}$, $\bm Z^{\mathbb K,N-1}$, $\bm Z^{\mathbb F,N-1}$, $N$, $A$, $M$.\\
		\midrule
		Initialize $\bm{\hat x}_m^{\mathbb K}(0|0)=\bm{\bar x}_m^{\mathbb K}(0)$, $\bm P_m(0|0)=\bm{\bar P}(0)$ and $\bm x_m^{\mathbb F}(\Xi)=\bm {\bar x}_m^{\mathbb F}$.\\
		\sf{For $t=1$ to $N-1$ } \\
		\hspace{0.4cm} 1, Predict the kinematic states and covariance according to \eqref{eq:xpredic}-\eqref{eq:Spredic}.\\
		\hspace{0.4cm} \sf{If $t \leq A-1$} \\
		\hspace{0.4cm} 2, Compute the associated measurement $\bm{\tilde z}_m^{\mathbb K}(t)$ for each track where the \\ 
		\hspace{0.8cm} association probability $\beta_{m,r}^{\mathbb K,\mathbb F}(t)$ is given by \eqref{eq:beta2}.\\
		\hspace{0.4cm} \sf{Else} \\
		\hspace{0.4cm} 3, Compute the associated measurement $\bm{\tilde z}_m^{\mathbb K}(t)$ for each track where the\\ 
		\hspace{0.8cm} association probability $\beta_{m,r}^{\mathbb K}(t)$ is given by \eqref{eq:beta}.\\
		\hspace{0.4cm} \sf{End if.}\\
		\hspace{0.4cm} 4, Update the tracks according to \eqref{eq:xupda}-\eqref{eq:Pupda2}.\\
		\sf{End for.} \\
		\hspace{0.4cm} 5, Estimate the feature signal $\bm {\hat x}_m^{\mathbb F}$ by iterating \eqref{eq:UPDAUpsilon}, \eqref{eq:UPDAtheta}-\eqref{eq:UPDAPhi}, \eqref{eq:UPDAx3}-\eqref{eq:UPDAx6}.\\
		\hspace{0.4cm} 6, Estimate the frequency of the feature by solving $\gamma_m-|\langle \bm{\hat q}_m, \bm a(f,0) \rangle|^2=0$\\
		\hspace{0.8cm} where $\bm {\hat q}_m$ is the dual solution of the optimization.\\
		\sf{For $t=1$ to $N-A-1$ }\\
		\hspace{0.4cm} 7, Predict the kinematic states and covariance according to \eqref{eq:xpredic}-\eqref{eq:Spredic}.\\
		\hspace{0.4cm} 8, Compute the associated measurement $\bm{\tilde z}_m^{\mathbb K}(t)$ for each track where the \\ 
		\hspace{0.8cm} association probability $\beta_{m,r}^{\mathbb K,\mathbb F}(t)$ is given by \eqref{eq:beta2}.\\
		\hspace{0.4cm} 9, Update the tracks according to \eqref{eq:xupda}-\eqref{eq:Pupda2}.\\
		\sf{End for.} \\
		\hspace{0.4cm} 10, $\bm{\bar x}_m^{\mathbb K}(0)=\bm{\hat x}_m^{\mathbb K}(N-A-1|N-A-1)$, $\bm{\bar P}_m(0)=\bm{P}_m(N-A-1|N-A-1)$.\\
		\hspace{0.4cm} 11, $\bm {\bar x}_m^{\mathbb F}=[\hat x_m^{\mathbb F}(N-A-1),\hat x_m^{\mathbb F}(N-A),...,\hat x_m^{\mathbb F}(N-1)]^T$.\\
		\midrule
		Return $\bm {\bar x}_m^{\mathbb K}(0)$, $\bm{\bar P}_m(0)$, $\bm {\bar x}_m^{\mathbb F}$, $\bm{\hat x}_m^{\mathbb K}(t|t)$ and $\bm{P}_m(t|t)$ for $m=1,2,...,M$ and\\ 
		$t=0,1,...,N-A-1$, then slide the batch forward.\\
	\end{tabular}
\end{algorithm}

\section{Simulation Results}
\label{sec:sim}
\subsection{Simulation Setup}

\begin{figure}[htbp]
	\centering
	\subfloat{\includegraphics[width=3in]{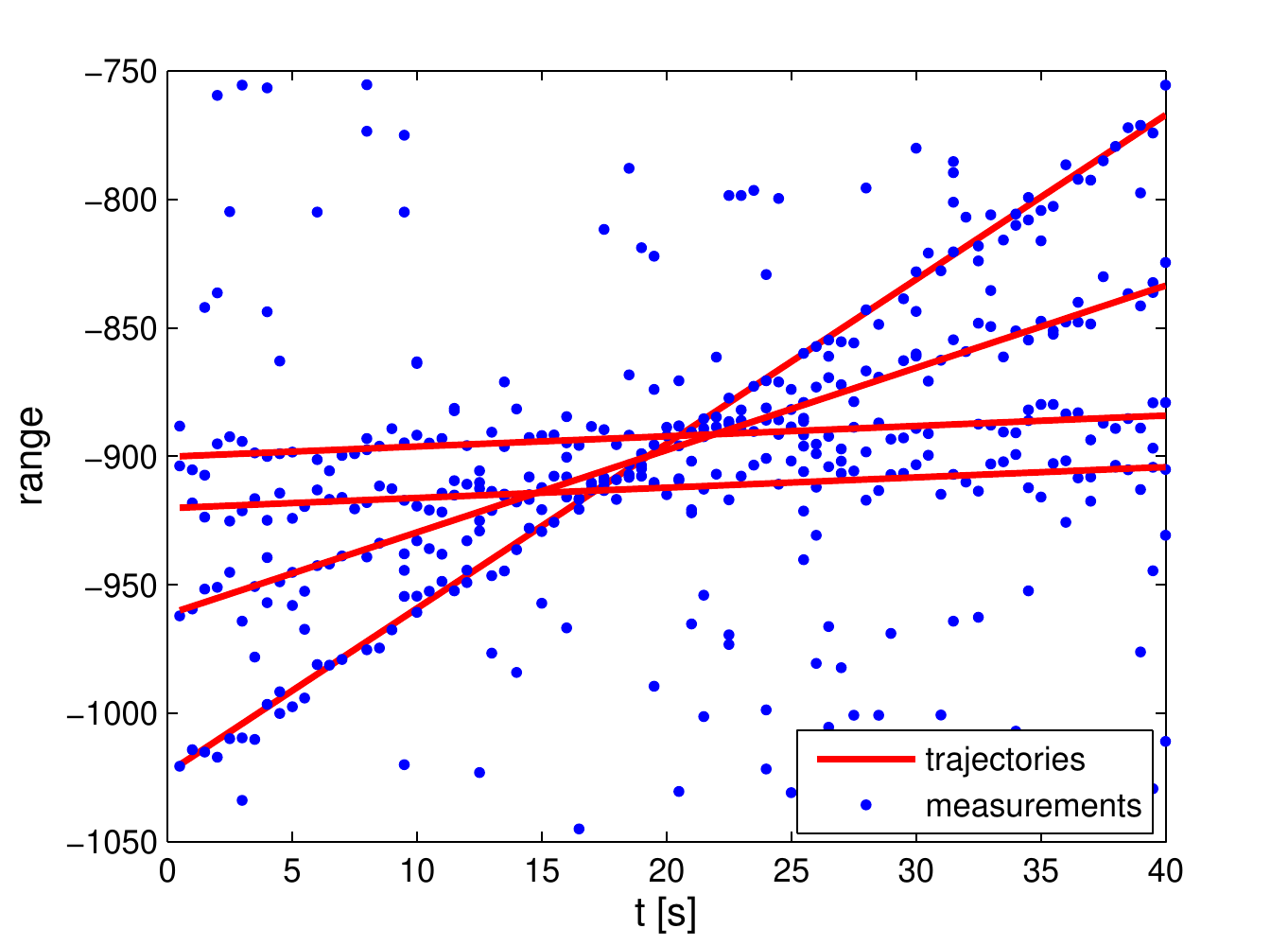}}
	\caption{The tracking scenario in the simulations.}
	\label{fig:scenario}
\end{figure}

To demonstrate the performance of the proposed algorithm, we perform experiments with the model of example given in Section II-C. There are four targets in the simulation. The initial states of the targets are $\bm x_1^{\mathbb K}(0)=[-1020,3.2]^T$, $\bm x_2^{\mathbb K}(0)=[-960,1.6]^T$, $\bm x_3^{\mathbb K}(0)=[-920,0.2]^T$ and $\bm x_4^{\mathbb K}(0)=[-900,0.2]^T$ respectively. Fig. \ref{fig:scenario} shows the trajectories of our scenario. As can be seen from the figure, targets 3 and 4 are moving in parallel, while the other two targets intersect with them.

The targets are tracked in range; that is, with reference to Eq. (\ref{eq:x}) (\ref{eq:y}), we have
\begin{equation*}
\bm F_m = \left[ {\begin{array}{*{20}{c}}
	1&{\Delta t} \\
	0&1	\end{array}} \right] \ \ \ \ \ \ \ 
\bm Q_m = \kappa _m^2\left[ {\begin{array}{*{20}{c}}
	{\frac{{\Delta {t^4}}}{4}}&{\frac{{\Delta {t^3}}}{2}} \\
	{\frac{{\Delta {t^3}}}{2}}&{\Delta {t^2}} \end{array}} \right]
\end{equation*}
with $80$ snapshots of measurements and the sampling interval $\Delta t = 0.5$s. Computing the derivative of the velocity, the acceleration of the $m$-th target is $-\varrho_m (2 \pi f_m)^2 {\rm sin}(2 \pi f_m (t-1) \Delta t)$. Hence, the acceleration varies from $-\varrho_m (2 \pi f_m)^2$ to $\varrho_m (2 \pi f_m)^2$. We assume the acceleration to be uniformly distributed, then the variance of the process noise should be 
\begin{eqnarray}
\label{eq:sigmap}
\kappa _m^2 = \frac{(2 \varrho_m (2 \pi f_m)^2)^2}{12} = \frac{\varrho_m^2 (2 \pi f_m)^4}{3}. 
\end{eqnarray}

We compare the performance of the FA-NN-JPDAF with that of the original NN-JPDAF and an augmented NN-JPDAF algorithm. For all the algorithms, the nonunity probability parameter is set as $B=0$ and the threshold for stopping the association is set as $\eta=0.15$, which is the same as the setting in \cite{muvsicki2007efficient}. The augmented NN-JPDAF algorithm is the NN-JPDAF with a state vector augmented with the feature and its changing rate, i.e., $\bm x_m^{\mathbb A} (t) = [x_m^{\mathbb K} (t,1), x_m^{\mathbb K} (t,2), x_m^{\mathbb F} (t), \dot x_m^{\mathbb F} (t)]^T$ with $\dot x_m^{\mathbb F} (t)$ denoting the changing rate of feature. The state transition process of the feature is modeled by the constant velocity (CV) model.

Some other parameters of the simulations are set as follows.

1, Targets are detected with a probability of $P_d=0.9$, and there are clutters whose density is $\mu$ at each time, generated uniformly in a viewing interval bounded by $1 \times {10^3}$. In our simulations, we set $\mu = 5 \times {10^{-3}}$, so the average number of clutters is 5 in each time. The amplitude of the clutters is uniformly distributed in an interval of $[0.5,1.5]$ and the phase of the clutter is uniformly distributed within $[0, 2 \pi]$.

2, The kinematic measurements are presented to the processor as ranges, where the measurement covariance $\bm R_m(t)$ is a scalar. We set $\bm R_m(t)=25$ for our simulations.

3, In each of the cases, simulations are run over 1000 random realizations. The tracks are initialized by single state whose distribution is assumed to be $p(\bm x_m^{\mathbb K}(0|0)) = N(\bm x_m^{\mathbb K}(0), \bm P_m(0|0))$. Here variable $\bm x_m^{\mathbb K}(0)$ represents the true state of the target at the first time step. The covariance matrix is initialized as $\bm P_m(0|0) = \text{diag}([10,10])$.

4, A track is declared to be lost at time $t$ if the position root-mean-squared-error (RMSE) is larger than $10\sigma _m$ or if the normalized estimation error squared (NEES) is larger than 20. The NEES of the $m$-th target is calculated as
\begin{eqnarray}
\label{eq:NEES}
\text{NEES}_m(t)= \left( \bm x_m^{\mathbb K}(t|t) - \bm x_m^{\mathbb K}(t) \right)^T \bm P_m^{-1}(t|t) \left( \bm x_m^{\mathbb K}(t|t) - \bm x_m^{\mathbb K}(t) \right) ,
\end{eqnarray}
in each time. The track continuity (percentage of tracks that are not lost) is calculated for the evaluation of the algorithms.

5, To check the accuracy of the tracking, we also calculate the position RMSE of the tracks that are not lost in the simulation. The proposed algorithm not only estimates the kinematic states of the targets, but also estimates the feature signal of the targets. So we also examine the accuracy of the feature extraction for the tracks that are not lost.

6, The frequencies of viberation are set as 0.6, 0.6, 0.8 and 0.8Hz for the four targets, respectively. According to \eqref{eq:sigmap}, the process noise depends on $f_m$ and $\varrho_m$. For simplicity, we set the magnitude of the vibration as 0.0244, 0.0244, 0.0137 and 0.0137 for the four targets, respectively. As a result, the standard deviation of the process noise is $\kappa_m=0.2$ for all targets.

7, The proposed FA-NN-JPDAF algorithm estimates the feature signal and its frequency, then the vibration frequency can be estimated by the distance between the strongest frequency component and the second strongest frequency component. To check the accuracy of the frequency estimation, we calculate the RMSE of the vibration frequency for the tracks that are not lost in the simulation.

8, For the proposed FA-NN-JPDAF algorithm, the parameters for feature extraction are set as $\gamma_m = \sigma_m \sqrt{N\log(N)}$, $\lambda_m = \frac{\gamma_m}{\sqrt{\alpha_m}}$, $\zeta_m=1$ and $\rho=0.1$. In the re-filtering step, we set $\tilde \sigma_m = \sqrt{10}\sigma_m$ for $m=1,2,...,M$.

9, For the augmented NN-JPDAF, the feature signal $x_m^{\mathbb F} (t)$ and its changing rate $\dot x_m^{\mathbb F} (t)$ are initialized by the true value. The kinematic states are initialized in the same way as that of the other two algorithms. The covariance matrix is initialized by $\bm P_m(0|0) = \text{diag}([10,10,10,10])$. The variance of feature process noise is set as $\frac{|\ddot x_{\max}^{\mathbb F}|^2}{3}$ in the filtering, where $|\ddot x_{\max}^{\mathbb F}|$ is the largest acceleration of feature signal.

\subsection{Tracking Performance}
\label{sec:comparison}

\begin{figure}[htbp]
	\centering
	\subfloat[][]{\includegraphics[width=3.2in]{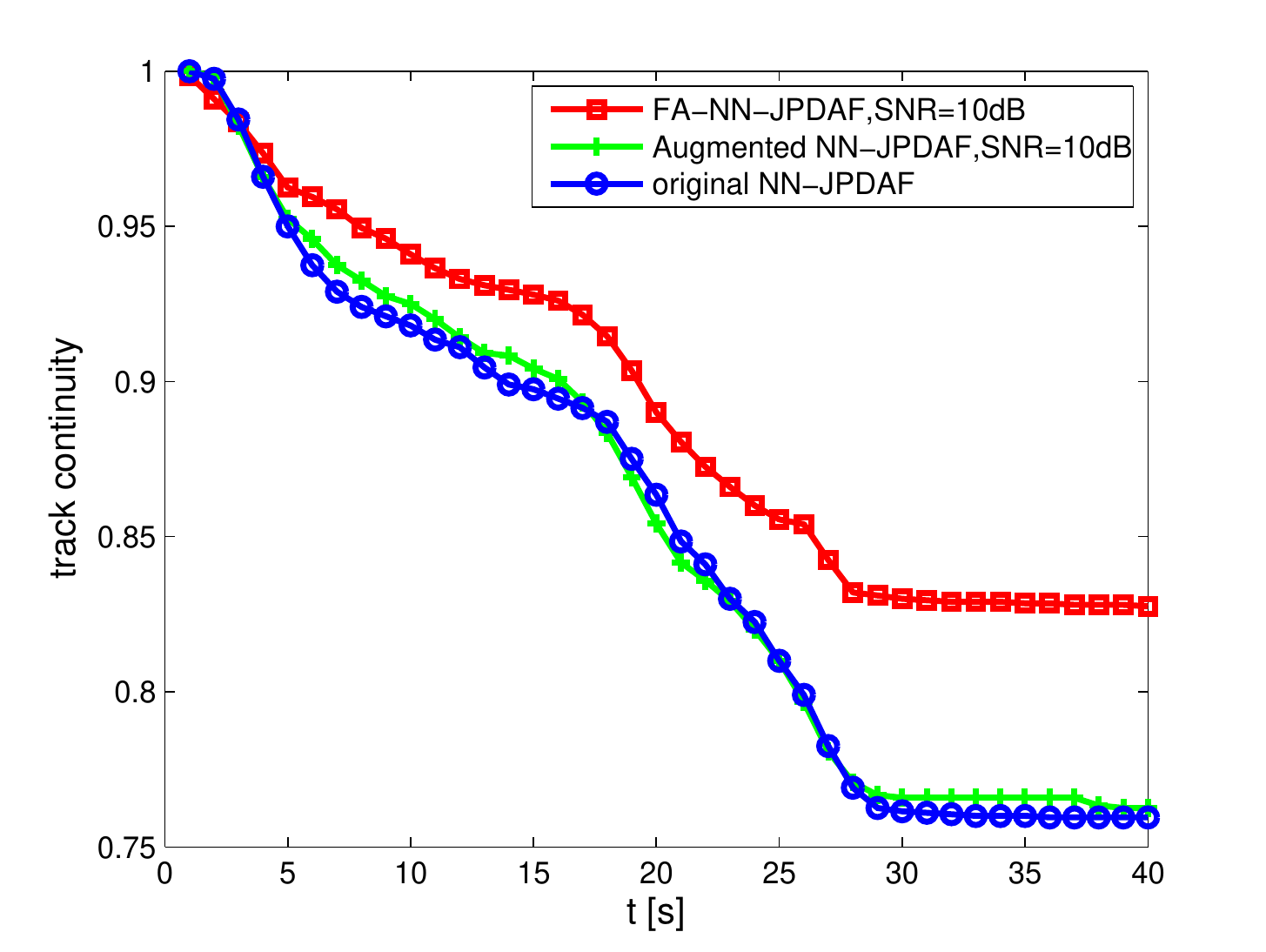}}
	\subfloat[][]{\includegraphics[width=3.2in]{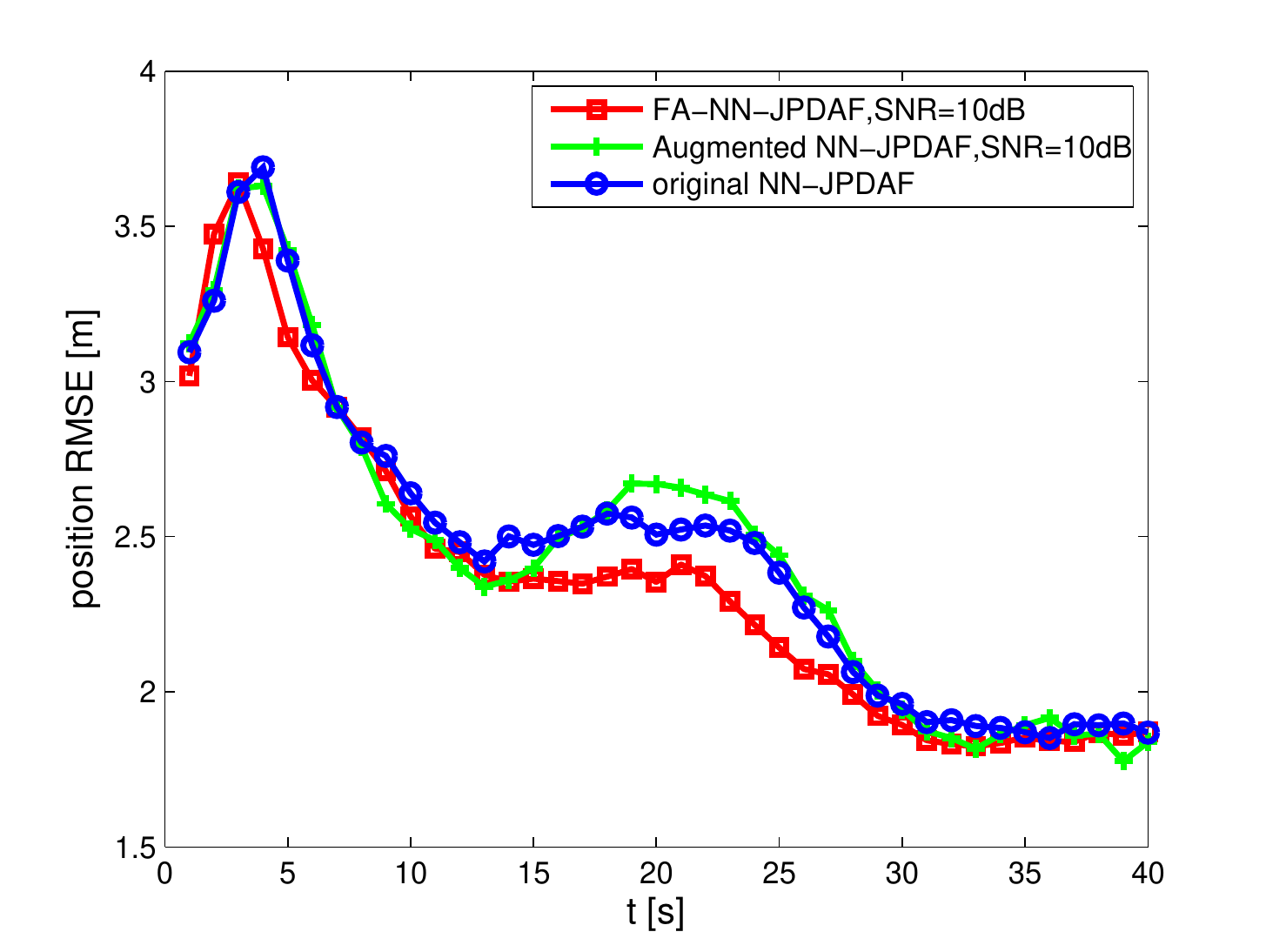}}
	\caption{Performance comparison: (a) track continuity, (b) position RMSE.}
	\label{fig:compare}
\end{figure}

In Fig. \ref{fig:compare}, the track continuity and position RMSE of the algorithms are plotted versus time, with the SNR of feature measurements set as 10dB. For the FA-NN-JPDAF, the length of the sliding batch is 32, with a skip parameter of 16 time steps. Hence, the overlap between adjacent batches is 16 time steps. Missed detections, false alarms and interference brought by multiple targets are the cause of the track losses. At the beginning of the track, the position RMSE grows with time. As time goes by, some tracks are lost while the other tracks become stable. Consequently, the position RMSE falls. As can be seen from the figure, the proposed algorithm can improve the performance of data association through the feature information, which results in better performance. More specifically, the improvement in the track-continuity is 6.8\% and the position RMSE of tracking is 3.0\% less than that of the original NN-JPDAF. The augmented NN-JPDAF algorithm cannot take advantage of the feature information and its performance is almost the same as that of the original NN-JPDAF.

Then we illustrate the performance of the proposed algorithm under different feature SNR. The feature SNR varies from 0dB to 20dB, with a step size of 2dB. The performance gain brought by the feature can be obtained by comparing the FA-NN-JPDAF and the original NN-JPDAF. We also compare the performance of the FA-NN-JPDAF with different overlap parameters. The number of the overlapped time steps are set as $A=16$ and $A=1$, respectively. In the second case, the overlap is very small, so the term $\frac{\zeta_m}{2}\|\bm x_m^{\mathbb F}(\Xi)-\bm {\bar x}_m^{\mathbb F}\|_2^2$ provides little help to feature estimation.

\begin{figure}[htbp]
	\centering
	\subfloat[][]{\includegraphics[width=3.2in]{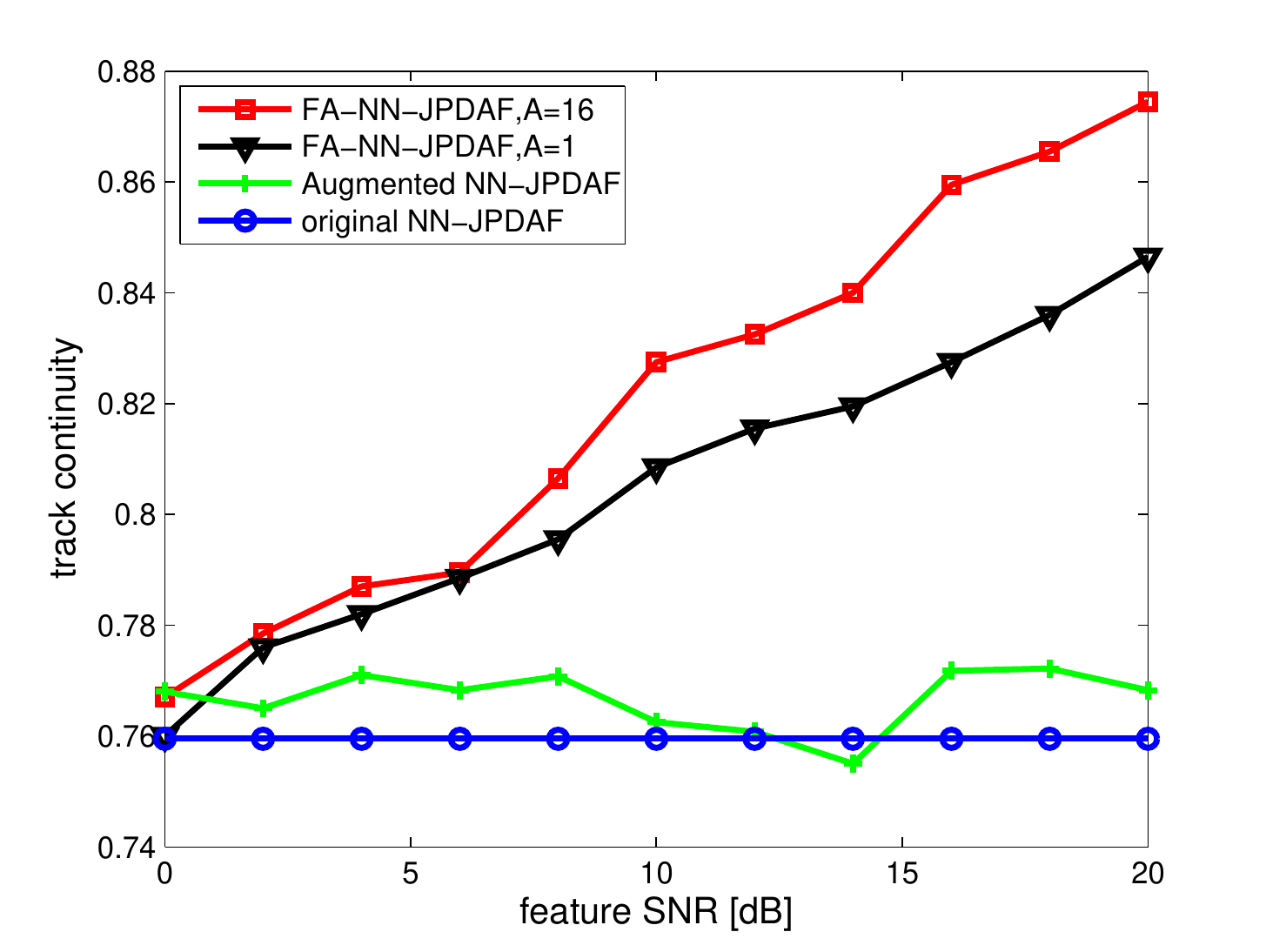}}
	\subfloat[][]{\includegraphics[width=3.2in]{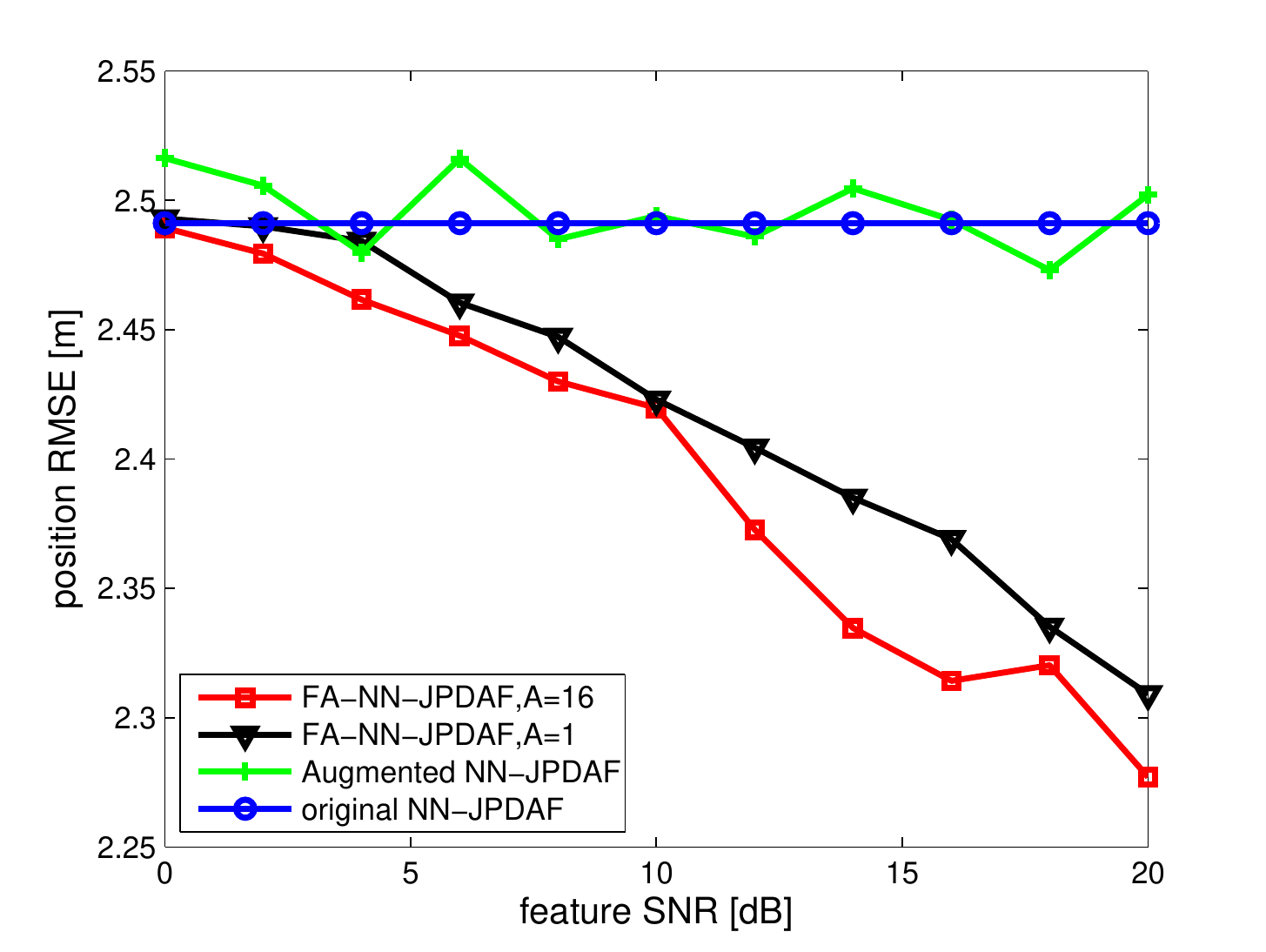}}
		
	\caption{Tracking performance for different feature SNR: (a) correct rate of tracking, (b) accuracy of position estimation.}
	\label{fig:compareSNR}
\end{figure}

In Fig. \ref{fig:compareSNR}, the correct rate of tracking and the accuracy of position filtering are plotted against the feature SNR. The proposed algorithm provides better performance than the other algorithms. When the feature SNR is 20dB and $A=16$, the correct tracking rate of FA-NN-JPDAF is over 11.5\% better than that of the original NN-JPDAF. As the feature SNR decreases, the error in feature estimation increases and the improvement brought by the feature information decreases as a result. When the feature SNR is 0dB, the performance gain brought by the feature information is small. In the case of $A=1$, the accuracy of feature estimation is worse, thereby reducing the performance gain. Specifically, when the feature SNR is 20dB, the performance gain in the correct tracking rate is 8.7\% over the original NN-JPDAF.

\begin{figure}[htbp]
	\centering
	
	\subfloat[]{\includegraphics[width=3.2in]{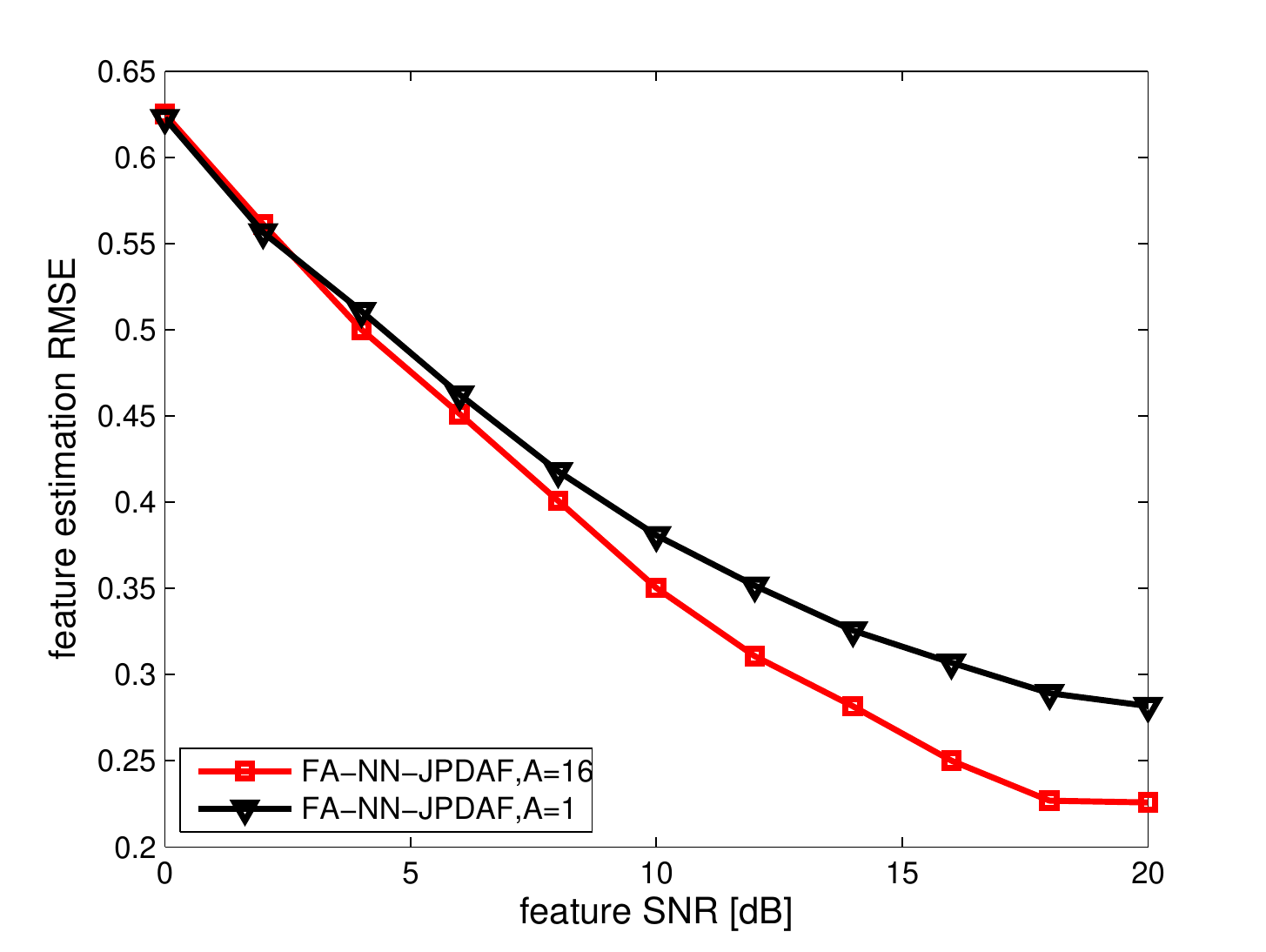}}
	\subfloat[]{\includegraphics[width=3.2in]{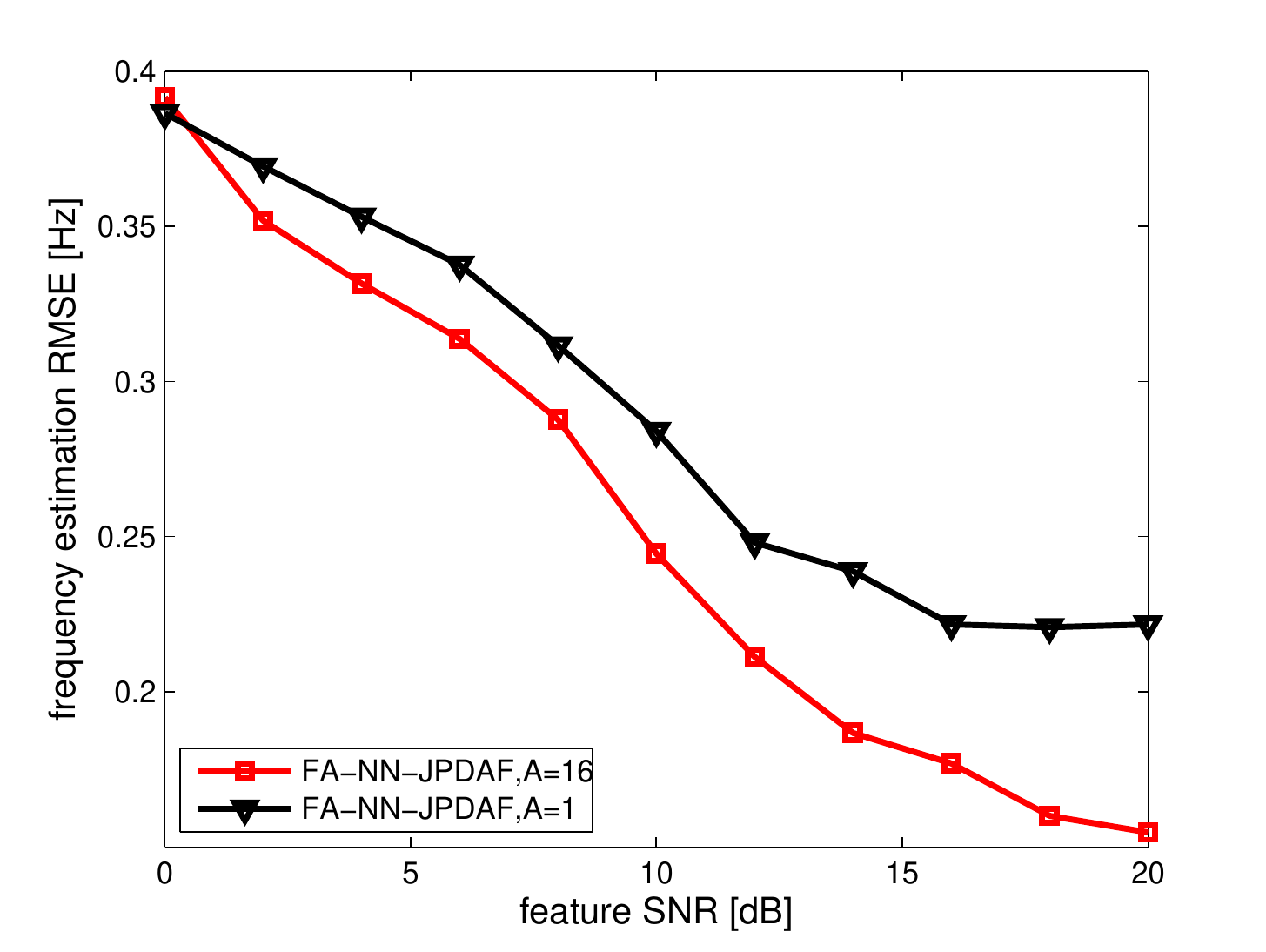}}
	
	\caption{Performance feature extraction: (a) accuracy of feature estimation, (b) accuracy of vibration frequency estimation.}
	\label{fig:featureSNR}
\end{figure}

Fig. \ref{fig:featureSNR} shows the feature estimation performance of the proposed algorithm. The RMSE of feature estimation and vibration frequency estimation are plotted against the feature SNR, respectively. As can be seen from the figure, the accuracy of the feature estimation increases as the feature SNR grows. Moreover, FA-NN-JPDAF with larger $A$ performs better, indicating that the information from the previous batch leads to better accuracy in feature estimation. It is also worth noting that the improvement is at the expense of some computational cost, so it will reduce the efficiency of the algorithm.

We offer several further remarks on our simulations:

\begin{enumerate}
	\item One advantage of the proposed algorithm is that it can use the phase information in the signal to improve the tracking performance. In many applications such as tracking radar, the received signal is complex-valued. Phase information is important because it is usually hard to discrimate the targets just by their amplitudes. However, many existing feature-aided tracking algorithms only use the amplitude information of the signals, and the phase information is neglected \cite{willett2001integration,song2009probabilistic,slocumb2005multiple}, so they cannot work if the targets have the same amplitude.
	
	\item The proposed algorithm does not require the frequencies of multiple targets to be different. In our simulations, target 1 and target 2 have the same vibration frequency while target 3 and target 4 have the same vibration frequency. Since the phase of the signal are different, the feature of the different targets are different in each time step, which helps us to improve the tracking performance.
	
	\item For the convenience of our analysis, the amplitude of the signal is assumed to be invariant over time and the detection probability is assumed to be invariant under different feature SNR in the simulations. Although it may not be true in practice, the main conclusions in the paper are not affected. In a more realistic setting, based on the estimation of signal strength, the detection threshold can also be optimized dynamically in the tracking. However, as our purpose in this paper is to provide an algorithm for a general tracking scenario rather than just for radar tracking, more detailed discussions will be left for the future work.
	
\end{enumerate}

\section{Conclusions}
\label{sec:conclu}
In this paper, a feature-aided NN-JPDAF algorithm is proposed for multiple target tracking. The proposed algorithm can extract the rapidly varying feature of the targets from the measurements when missed detections and mis-associations exist. The extracted feature is then used to aid the data association. Simulation results show that the proposed algorithm provides better tracking performance compared to the original NN-JPDAF and an augmented NN-JPDAF algorithm. The underlying ideas of this paper can also be applied to the other tracking algorithms such as MHT and PMHT, which is under investigation. Future work should also include the incorporation of target maneuver and multiple sensors.

\appendix
\subsection{Proof of Lemma 1}
Since strong duality holds, we have
\begin{eqnarray}
\label{eq:left}
&&\gamma_m\|\bm {\hat x}_m^{\mathbb F} \|_{\cal A}+\lambda_m \|\bm {\hat e}_m^{\mathbb F}\|_1 + \frac{1}{2} \|\bm{\tilde z}_m^{\mathbb F}-\bm{\hat x}_m^{\mathbb F}(\Omega_m)-\bm{\hat e}_m^{\mathbb F}\|_2^2 \nonumber \\
&&= {\left\langle {{\bm{\hat q}_m}({\Omega _m}),\bm{\tilde z}_m^{\mathbb F}} \right\rangle _{\mathbb R}} - \frac{1}{2}\left\| {\bm{\hat q}_m} \right\|_2^2 , \nonumber \\
&&= \langle \bm{\hat q}_m(\Omega_m) , \bm{\hat x}_m^{\mathbb F}(\Omega_m) \rangle_{\mathbb R} + \langle \bm{\hat q}_m(\Omega_m) , \bm{\hat e}_m^{\mathbb F} \rangle_{\mathbb R}  \nonumber \\ 
&& ~~ + \langle \bm{\hat q}_m(\Omega_m) , \bm{\tilde z}_m^{\mathbb F} - \bm{\hat x}_m^{\mathbb F}(\Omega_m) - \bm{\hat e}_m^{\mathbb F} \rangle_{\mathbb R} - \frac{1}{2}\left\| {\bm{\hat q}_m} \right\|_2^2.
\end{eqnarray}
Now we want to prove $\gamma_m\|\bm {\hat x}_m^{\mathbb F} \|_{\cal A}+\lambda_m \|\bm {\hat e}_m^{\mathbb F}\|_1 = \langle \bm{\hat q}_m , \bm{\hat x}_m^{\mathbb F} \rangle_{\mathbb R} + \langle \bm{\hat q}_m(\Omega_m) , \bm{\hat e}_m^{\mathbb F} \rangle_{\mathbb R}$ which can be achieved in two steps. Based on \eqref{eq:left}, we firstly have
\begin{eqnarray}
\label{eq:left2}
\gamma_m\|\bm {\hat x}_m^{\mathbb F} \|_{\cal A}+\lambda_m \|\bm {\hat e}_m^{\mathbb F}\|_1 \nonumber &=& \langle \bm{\hat q}_m(\Omega_m) , \bm{\hat x}_m^{\mathbb F}(\Omega_m) \rangle_{\mathbb R} + \langle \bm{\hat q}_m(\Omega_m) , \bm{\hat e}_m^{\mathbb F} \rangle_{\mathbb R}  \nonumber \\ 
&& - \frac{1}{2} \|\bm{\tilde z}_m^{\mathbb F}-\bm{\hat x}_m^{\mathbb F}(\Omega_m)-\bm{\hat e}_m^{\mathbb F}-\bm{\hat q}_m(\Omega_m) \|_2^2, \nonumber \\
&\leq& \langle \bm{\hat q}_m(\Omega_m) , \bm{\hat x}_m^{\mathbb F}(\Omega_m) \rangle_{\mathbb R} + \langle \bm{\hat q}_m(\Omega_m) , \bm{\hat e}_m^{\mathbb F} \rangle_{\mathbb R} , \nonumber \\
&=& \langle \bm{\hat q}_m , \bm{\hat x}_m^{\mathbb F} \rangle_{\mathbb R} + \langle \bm{\hat q}_m(\Omega_m) , \bm{\hat e}_m^{\mathbb F} \rangle_{\mathbb R}.
\end{eqnarray}
Then, using the fact that $\|\bm{\hat q}_m \|_{\cal A}^* \leq \gamma_m$ and $\|\bm{\hat q}_m \|_\infty \leq \lambda_m$, we have
\begin{eqnarray}
\label{eq:right}
\gamma_m \|\bm {\hat x}_m^{\mathbb F} \|_{\cal A}+\lambda_m \|\bm {\hat e}_m^{\mathbb F}\|_1 &\geq& \| \bm{\hat q}_m \|_{\cal A}^* \| \bm{\hat x}_m^{\mathbb F} \|_{\cal A} +  \| \bm{\hat q}_m \|_\infty \| \bm{\hat e}_m^{\mathbb F} \|_1, \nonumber \\
&\geq& \langle \bm{\hat q}_m , \bm{\hat x}_m^{\mathbb F} \rangle_{\mathbb R} + \langle \bm{\hat q}_m(\Omega_m) , \bm{\hat e}_m^{\mathbb F} \rangle_{\mathbb R},
\end{eqnarray}
where the second inequality is a result of the H\"{o}lder's inequality. Combining \eqref{eq:left2} and \eqref{eq:right} leads to $\gamma_m \|\bm {\hat x}_m^{\mathbb F} \|_{\cal A}+\lambda_m \|\bm {\hat e}_m^{\mathbb F}\|_1=\langle \bm{\hat q}_m , \bm{\hat x}_m^{\mathbb F} \rangle_{\mathbb R} + \langle \bm{\hat q}_m(\Omega_m) , \bm{\hat e}_m^{\mathbb F} \rangle_{\mathbb R}$. Applying the fact that $\langle \bm{\hat q}_m , \bm{\hat x}_m^{\mathbb F} \rangle_{\mathbb R} \leq \gamma_m \|\bm {\hat x}_m^{\mathbb F} \|_{\cal A}$ and $\langle \bm{\hat q}_m(\Omega_m) , \bm{\hat e}_m^{\mathbb F} \rangle_{\mathbb R} \leq \lambda_m \|\bm {\hat e}_m^{\mathbb F}\|_1$, we have $\langle \bm{\hat q}_m , \bm{\hat x}_m^{\mathbb F} \rangle_{\mathbb R} = \gamma_m \|\bm {\hat x}_m^{\mathbb F} \|_{\cal A}$ and $\langle \bm{\hat q}_m(\Omega_m) , \bm{\hat e}_m^{\mathbb F} \rangle_{\mathbb R} = \lambda_m \|\bm {\hat e}_m^{\mathbb F}\|_1$. This is possible only if \eqref{eq:dualc} and \eqref{eq:dualc2} hold.

\bibliographystyle{IEEEtran}
\bibliography{database}
\end{document}